\documentclass[amsmath,amsfonts,amssymb,aps,11pt]{revtex4}
\usepackage{graphicx}
\usepackage{srcltx}

\begin{document}

%\def\lsim{\mathrel{\raise.3ex\hbox{$<$\kern-.75em\lower1ex\hbox{$\sim$}}} }
%\def\gsim{\mathrel{\raise.3ex\hbox{$>$\kern-.75em\lower1ex\hbox{$\sim$}}} }
%\newcommand{\Rbs}{\mbox{${{\scriptstyle \not}{\scriptscriptstyle R}}$}}

%%%%%%%%%%%%%%%%%%%%%%%%%%%%%% User specified LaTeX commands.
% definition to produce a "less than or similar to" symbol
\def\lsim{~\rlap{$<$}{\lower 1.0ex\hbox{$\sim$}}\;}
% definition to produce a "greater than or similar to" symbol
\def\gsim{~\rlap{$>$}{\lower 1.0ex\hbox{$\sim$}}\;}
\def\pslash{\displaystyle{\not}p}
\def\kslash{\displaystyle{\not}k}
\def\qslash{\displaystyle{\not}q}
%%%%%%%%%%%%%%%%%%%%%%%%%%%%%%%%%%%%%%%%%%%%%%%%%%%%%%%

\title{Probing Majorana neutrinos in rare $K$ and $D,~D_s,B,B_c$ meson decays}

\thispagestyle{empty}
%\begin{titlepage}
\author{ G. Cvetic$^{1,}$\footnote{E-mail: gorazd.cvetic@usm.cl},
Claudio Dib$^{1,}$\footnote{E-mail: claudio.dib@usm.cl},
 Sin Kyu Kang$^{2,}$\footnote{E-mail:
        skkang@snut.ac.kr},~~ C. S. Kim$^{3,}$\footnote{E-mail:
        cskim@yonsei.ac.kr,~~~Corresponding Author} }
%\thanks{ E-mail : skkang1@sogang.ac.kr}}
%\date{\today}
\affiliation{ $^{1}$ Centro Cientifico y Tecnologico de Valparaiso
and  Department of Physics, Universidad Tecnica Federico Santa
Maria, Valparaiso, Chile \\
             $ ^{2}$ School of Liberal Arts, Seoul National University of Technology,
                     Seoul 121-742, Korea \\
              $ ^{3}$ Dept. of Physics and IPAP, Yonsei University, Seoul 120-749, Korea }
%\end{titlepage}

\pacs{}
\begin{abstract}
%\hs
\noindent We study lepton number violating decays of charged $K$,
$D$, $D_s$, $B$ and $B_c$ mesons of the form $M^+\to {M'}^-\ell^+\ell^+$,
induced by the existence of Majorana neutrinos. These  processes
provide information  complementary to neutrinoless double nuclear
beta decays, and are sensitive to neutrino masses and lepton
mixing. We explore neutrino mass ranges $m_N$ from below 1 eV to
several hundred GeV. We find that in many cases the branching ratios are
prohibitively small, however in the intermediate range $m_\pi < m_N < m_{B_c}$,
in specific channels and for specific neutrino masses, the
branching ratios can be at the reach of high luminosity
experiments like those at the LHC-$b$ and future Super flavor-factories,
and can provide bounds on the
lepton mixing parameters.

\end{abstract}
\maketitle \thispagestyle{empty}

%%%%%%%%%%%%%%%%%%%%%%%%%%%%%%%%%%%%%
%
%                         INTRODUCTION
%
%%%%%%%%%%%%%%%%%%%%%%%%%%%%%%%%%%%%%
\section{Introduction}

One of the outstanding issues in neutrino physics today is to
clarify the  Dirac or Majorana character of neutrino masses.
The discovery of neutrino oscillations indicates that neutrinos
are massive particles with masses likely to be much smaller than
those of charged fermions \cite{SuperK}.
This fact provides an important clue on the existence of a more
fundamental physics  underlying the standard model (SM) of
particle physics, because neutrinos are naturally massless in the
SM.
Although the experimental results on neutrino oscillations can
determine the neutrino mixing parameters and their squared mass
differences, the absolute magnitudes of the masses as well as
their origin remain unknown and constitute fundamental open
questions in neutrino physics.
Many experiments have been set to search for the absolute
magnitude of neutrino masses.
Direct methods to determine the mass of the electron neutrino use
the endpoint of the electron spectrum in beta decays.
The most sensitive of these experiments uses
Tritium \cite{Absolutemass}, setting the present upper bound
$m_{\nu_e}<2$ eV  \cite{PDG2008}, and the next experiment is expected to
reach a sensitivity of $0.2$~eV \cite{Katrin}.
Other experiments do direct searches for muon and tau neutrino
masses, setting the upper bounds $ m_{\nu_\mu} <190$ keV and
$m_{\nu_\tau} <18.2$ MeV respectively, at $90\%$ C.L.  \cite{PDG2008}.
To date, the most stringent bound on the sum of all light neutrino
masses is obtained from cosmological observations, given by
$\sum_{i} m_{\nu_i} <0.17$ eV at $95\%$ C.L., a figure which is,
to a certain extent, model dependent \cite{cosmology}.

If neutrinos are Dirac particles, they must have right-handed
electroweak singlet components in addition to the known left-handed
modes; in such case lepton number remains as a conserved quantity.
Alternatively, if neutrinos are Majorana particles, then a
neutrino is indistinguishable from its antiparticle and lepton
number would be violated by two units ($\Delta L=2$) in some
processes that involve neutrinos.
The experimental results to date are unable to distinguish between
these two alternatives.

There have been several attempts to determine a
Majorana nature of neutrinos by studying $\Delta L=2$
processes. The most prominent of these
processes are neutrinoless nuclear double beta decays ($0\nu \beta
\beta $), which have  been regarded as the most sensitive way to
look for lepton number violation (LNV) \cite{nndb1}.
The observation of $0\nu \beta \beta $ would indeed be very
important not only because it would  establish the existence of
LNV -- implying that neutrinos are Majorana
particles, but also because they would provide a scale for the
absolute magnitude of light neutrino masses, complementary to the
direct searches mentioned above: these nuclear processes are
proportional to the square of the \emph{effective} neutrino mass
$m_{ee}=|\sum_{i=1}^3U^2_{ei}m_i|$, with $m_i$ and $U_{ei}$ being
the individual neutrino masses and the $\nu_i - e$ mixing matrix
elements, respectively \cite{nndb2}.
However, it has long been recognized that, even though the
experiments are very sensitive, the extraction of the neutrino
mass scale and the Majorana nature of neutrinos from nuclear $0\nu
\beta \beta $ is a difficult task, because reliable information on
neutrino properties can be inferred only if the nuclear matrix
elements for $0\nu \beta \beta $ are calculated correctly. The
calculation of the nuclear matrix elements for $0\nu \beta \beta$,
usually performed within either the quasi-particle random phase
approximation  \cite{rpa} or the nuclear shell model  \cite{nsm}
or their variants, is known to be a complex task, sometimes with
large differences among the different  approaches
\cite{comparison}. Even in the most refined treatments, the
estimates of the nuclear matrix elements remain affected by
various large uncertainties \cite{nmeun}.

Another avenue to detect the Majorana nature of neutrinos is to study $\Delta L=2$
processes  in rare meson decays \cite{Kova1, Ali, deltal2}.
In this paper we study  $\Delta L = 2$ decays of heavy
charged mesons whose signals could be captured at high intensity
experiments such as LHC-$b$ and  future Super $B$-factories as well as
advanced $K$-factories.
The $\Delta L= 2$ processes we treat in this paper are rare
neutrinoless decays of heavy charged mesons into a lighter meson
and two charged leptons of the same sign \cite{Ali}.
These processes, just like neutrinoless nuclear double beta
decays, can occur only via Majorana neutrino exchange, and thus
their experimental observation could establish the Majorana
character of the neutrinos and the absolute scale of neutrino
masses in much the same way as in nuclear $0\nu \beta \beta $
decays, but there are some essential differences.
{}From a theoretical viewpoint, the uncertainties in meson decays
are much easier to handle than in nuclear $0\nu \beta \beta$ decays.
However, from the experimental viewpoint, the $\Delta L=2$ meson decay rates in the case
of standard neutrinos ($m_\nu < 2$ eV) are prohibitively small for
any experiment, while $0\nu \beta \beta$ decays are more
realistic options, due to their macroscopically large samples of
decaying nuclei. In contrast, for heavier, non-standard,
neutrinos, the meson decay rates are good alternatives to search for,
as they can be within reach of future experiments.

In this study it is important to distinguish between standard and
sterile neutrinos. From direct searches we know the standard
electron neutrino mass is below 2 eV \cite{cosmology}, and
neutrino oscillation experiments tell us that all three neutrino
masses differ from one another by much less than that value
\cite{neutrino-osc}.
Therefore all neutrinos with
masses above 2 eV are assumed to be non-standard. Since our work
is mainly relevant for neutrinos above this bound, in what follows
we will denote them generically by the letter $N$, instead of
$\nu$.

An important motivation to search for sterile (non-standard)
neutrinos with masses of the order of 1 MeV is that their
existence has nontrivial observable consequences for cosmology and
astrophysics. They are presumed to participate in big-bang
nucleosynthesis, supernovae explosions, large scale structure
formation and, in general, to be a component of the dark matter in
the universe \cite{sdm}. Thus, sterile neutrino masses and their
mixing with the standard neutrinos must be subjected to cosmological
and astrophysical bounds \cite{sb1}. There are also some
laboratory bounds coming from the fact that  sterile neutrinos
contribute via mixing with the standard neutrinos to various
processes which are forbidden in the SM. Those bounds turn out to
be much weaker than the cosmological and astrophysical bounds, but
useful in cases where the latter become inapplicable
\cite{sb2}.

We have separated the analysis into three different cases,
depending on the relevant neutrino mass range.
If the exchanged neutrino is much lighter than the energy scale in
the process, the amplitude of the decay rate is proportional to
the square of an effective electron-neutrino mass,
$m_{ee}^2=|\sum_{N}U^2_{eN}m_N|^2$, which is anticipated to be  of
the order $\sim 1$ eV$^2$  or less from current neutrino data and
cosmological observations  such as WMAP  \cite{WMAP}, if only
standard neutrinos are involved.
Instead, if the exchanged neutrino is much heavier than the
decaying meson, the decay rate is proportional to $|U_{N
\ell_1}U_{N \ell_2}/m_N|^2$, where $m_N$ and $U_{N \ell}$ are the
heavy neutrino mass and its mixing with the standard leptons,
respectively. In general, in this case $U_{N \ell}$ is small and
$m_N$ is large, so the factor constitutes a severe suppression to
the decay rate.
Finally, for the case of Majorana neutrinos with intermediate
masses between that of the initial and the final meson, the decay
rate is dominated by a resonantly enhanced $s$-channel
amplitude \cite{Kova1, Ali, Kova}, where the intermediate neutrino goes on
its mass shell.

In Section II, we describe the approximation methods for the
calculations of rare heavy meson decays of the form $M^+ \to
M^{\prime -} l_1^+ l_2^+$ (where $M$ and $M'$ are pseudoscalar
mesons). Here we are interested on $K^+,~D^+,~D_s^+,~B^+$ and
$B_c^+$ decays into $\pi^- \ell^+ \ell^+$, $K^- \ell^+ \ell^+$,
$D^- \ell^+ \ell^+$, $D_s^- \ell^+ \ell^+$ and $B^- \ell^+
\ell^+$, (where $\ell =$ $e$, $\mu$ or $\tau$), therefore, we will
denote the initial and final mesons generically by $M^+$ and
$M'^-$, respectively. We separate the analysis for the three cases
of light neutrinos $(m_N < m_{M'})$,  intermediate neutrinos
$(m_{M'} < m_N < m_M)$, and  heavy neutrinos $(m_N>m_M)$. We
include the results and discussions in each subsection.
In Section III  we summarize the results and state our
conclusions.

%%%%%%%%%%%%%%%%%%%%%%%%%%%%%%%%%%%%%
%
%                        SECTION 2
%
%%%%%%%%%%%%%%%%%%%%%%%%%%%%%%%%%%%%%
\section{Calculations of $M^+ \to M^{\prime -} l_1^+ l_2^+$}

We now describe our approximation methods for the
calculations of rare heavy meson decays of the form $M^+ \to
M^{\prime -} l_1^+ l_2^+$ (where $M$ and $M'$ are pseudoscalar
mesons) in all three neutrino mass ranges described above.
\begin{figure}[b]
\begin{minipage}[b]{.45\linewidth}
%\centering
 \includegraphics[width=\linewidth]{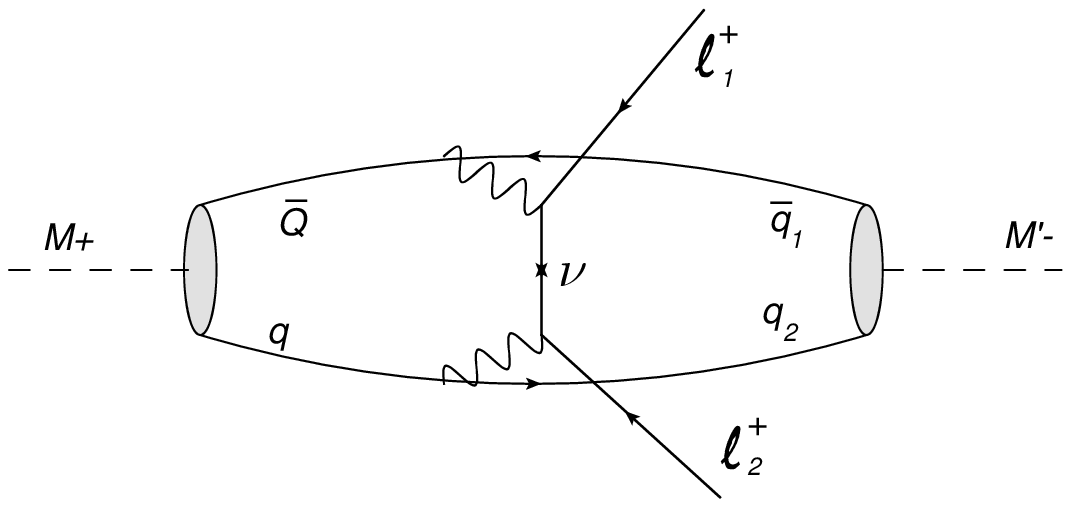}
\end{minipage}
\begin{minipage}[b]{.05\linewidth}
\hspace{1pt}
\end{minipage}
\begin{minipage}[b]{.45\linewidth}
\vspace{0pt}
 \includegraphics[width=\linewidth]{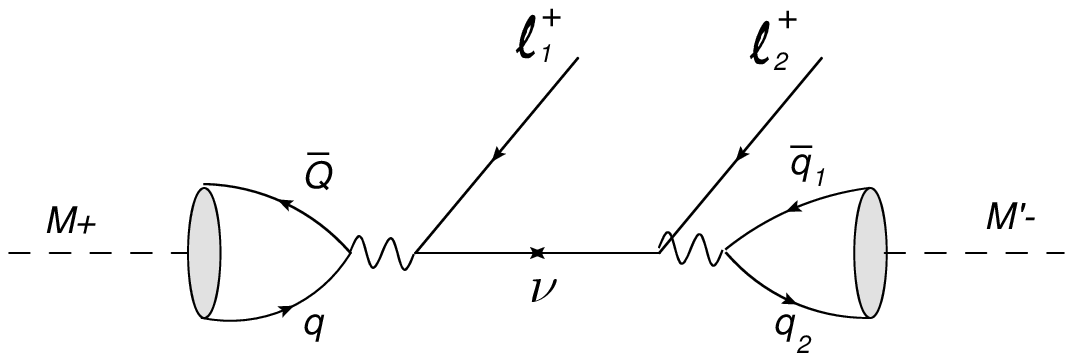}
% \vspace{0pt}
\end{minipage}
\vspace{-0.2cm} \caption{The $t$-type and $s$-type  weak amplitudes at the quark
level that enter in the process $M^+\to {M'}^- \ell_1^+ \ell_2^+$
(plus the same diagrams with leptons exchanged if they are
identical).} \label{fig1}
\end{figure}
At the quark level, the decay occurs via two types of amplitudes,
shown in Fig.~\ref{fig1}.
We find that in the case of light neutrinos ($m_N < m_\pi$), the
amplitude on the left in Fig.~\ref{fig1} (``$t$-type" diagram)
dominates due to long distance
contributions, and the decay rate becomes proportional to $m_{M}^7
\times m_N^2$. For this reason, only the decays of the heavier $B$
mesons are of any importance in this case.
In contrast, for intermediate neutrino masses ($m_{M'} < m_N <
m_M$), the diagram on the right in Fig.~\ref{fig1} (``$s$-type" diagram)
 dominates when the neutrino propagator becomes
resonant on its mass shell, in which case the decay rate turns out
to be less dependent of the neutrino mass, but very sensitive to
the mixing elements.
Finally,  for heavy neutrinos ($m_N > m_M$), both amplitudes in
Fig.~\ref{fig1} are comparable and the decay rate is $\propto
1/m_N^2$.

In Table \ref{table1} we list the numerical values of the input parameters
we use in our numerical estimates.

\begin{table}
\caption{Values of input parameters used in our calculations.
They correspond to the central values given in Ref.~\cite{PDG2008},
except for $f_B$ and $f_{B_s}$ which are taken from Ref.~\cite{CKWN},
and $V_{cs}$ which is calculated by imposing the unitarity constraint
on the CKM matrix.} \label{table1}
\begin{tabular*}{0.50\textwidth}{@{\extracolsep{\fill}} |ll||ll| }
\hline
 Parameter & Value  & Parameter & Value
\\
\hline
$f_\pi$ & 130.4 [MeV] & $V_{ud}$ & 0.9742
\\
$f_K$ & 155.5 [MeV]  & $V_{us}$ & 0.2255
\\
$f_{D^+}$ & 205.8 [MeV] & $V_{ub}$ & 0.0039
\\
$f_{D_s}$ & 273. [MeV] &  $V_{cd}$ & -0.230
\\
$f_B$ & 196. [MeV] & $V_{cs}$ & 0.950
\\
$f_{B_c}$ & 322. [MeV] &  $V_{cb}$ & 0.041
\\
\hline
\end{tabular*}
\end{table}

%%%%%%%%%%%%%%%%%%%%%%%%%%%%%%%%%%%%%
%
%                         SUBSECTION 2.1    LIGHT CASE
%
%%%%%%%%%%%%%%%%%%%%%%%%%%%%%%%%%%%%%

\subsection{ The case of light neutrinos ($m_N < m_\pi$)}

\begin{figure}[t]
  \centering
  \includegraphics[scale=0.8]{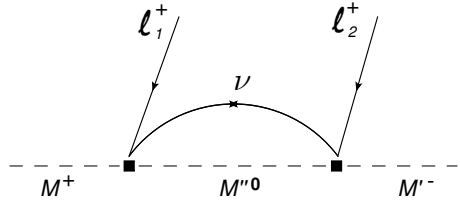}
  \caption{The main diagram in an effective meson theory for
$M^+\to M^{\prime -}\ell^+\ell^+$ (plus diagram with leptons
exchanged if they are identical), mediated by Majorana neutrinos,
when the neutrino is much lighter than the final meson. The
amplitude is estimated considering the intermediate state on its
mass shell.}
  \label{fig2}
\end{figure}

%A neutrinoless decay
    We find that a neutrinoless decay
like $B^+ \to D^- \ell^+ \ell^+$ with light
Majorana neutrinos in the intermediate state is dominated
    at the meson level
by the amplitude shown in Fig.~\ref{fig2},
    when the intermediate state goes on mass shell.
    This amplitude originates at the quark level from the $t$-type weak amplitude shown in Fig.~\ref{fig1}.
    We find the $s$-type amplitude shown in Fig.~\ref{fig1} to be subdominant, or at most comparable with
    the former. In this sense, our treatment differs from that of A.~Ali {\it et al.} \cite{Ali}, where the $s$-type amplitude is assumed to dominate \cite{Controversial}.
    However, since the rate in any case turns out to be too small for any foreseeable experiment, we will just do an order-of-magnitude estimate for it, calculating the absorptive part and assuming that
    the dispersive part is not much larger. The absorptive part of the amplitude is calculated by setting the intermediate particles on their mass shell and then integrating over their phase space:
\begin{equation}
{\cal M}_{\rm abs}(B^+\to D^- \ell^+ \ell^+) = \int d{\rm ps}_{D
N}\ A_{B\to D N \ell}\ A_{D N\to D\ell}
 \label{Mabs1}
\end{equation}
where  $A_{B\to D N\ell}$ and $A_{D N\to D\ell}$ are the
tree-level amplitudes for the respective sub-processes, and
 $d{\rm ps}_{D N}$ is the Lorentz-invariant phase space of the
intermediate $D$-$N$ pair, which in the rest frame of the pair is
$
d{\rm ps}_{D N} =\sum_{s} (1/16\pi^2) (|{\bf p}_N| / m_{D\ell})
d\Omega_N $. Here $\sum_s$ is the sum over the neutrino spins,
${\bf p}_N$ is the 3-momentum of the neutrino in the $D$-$N$ rest
frame, and $m_{D\ell}$ is the invariant mass of the pair.
In turn, the amplitudes of the weak sub-processes are:
\begin{eqnarray}
A_{B^+\to D^0 N\ell} &=& \frac{G_F}{\sqrt{2}} \ V_{cb}\  U_{N\ell} \ \langle
\bar D^0(p')|J^\mu(0)|B^+(p)\rangle
   \  \bar u_{N}(p_N) \gamma_\mu (1-\gamma_5) v_{\ell}(l_1)
\label{amplitudes1}\\
 A_{D^0 N\to D^-\ell} &=& \frac{G_F}{\sqrt{2}} \  V_{ud} \  U_{N\ell}\ \langle
D^-(p')|J^\mu(0)|\bar D^0(p)\rangle
   \  \bar v_{N}(p_N) \gamma_\mu (1-\gamma_5) v_{\ell}(l_2),\nonumber
\end{eqnarray}
where $V$ is the Cabbibo-Kobayashi-Maskawa (CKM) matrix for quark
mixing, and $U$ the Pontecorvo-Maki-Nakagawa-Sakata (PMNS) matrix
for lepton mixing. The hadronic matrix elements can be
parameterized in terms of phenomenological form factors $F^+(q^2)$
and $F^-(q^2)$ as
\begin{equation}
\langle \bar D^0(p')|J^\mu(0)|B^+(p)\rangle
=
F_{BD}^+(q^2)(p+p')^\mu + F_{BD}^-(q^2) (p-p')^\mu ,
\label{formfactors}
\end{equation}
and similarly for $\langle D^-(p')|J^\mu(0)|\bar D^0(p)\rangle $,
where $q$ is the corresponding 4-momentum transfer.  In our crude
estimate, we will neglect the $F^-$ form factors and assume the
$F^+$ to be constants of order unity over the kinematical range.

Now, in the product $A_{B\to D N\ell}\times A_{D N\to D\ell}$,
after summing over intermediate spin states, the two lepton lines
can be combined into a single one by using in $A_{D N\to D\ell}$ the
identity $\bar v_N\gamma^\mu(1-\gamma_5)v_{\ell} =
\bar u_{\bar\ell} \gamma^\mu(1+\gamma_5) u_N$.  Here $u_{\bar\ell}$ is a $u$-spinor for the
charged antilepton, and the neutrino is assumed to satisfy
the Majorana condition $u_{\bar N}  = \lambda_N \, u_N$ (where $\lambda_N$ is a phase).
The result is
then:
\begin{eqnarray}
{\cal M}_{\rm abs} &=& \frac{G_F^2}{2} V_{cb} V_{ud} \, U_{N\ell}^2 \,  \lambda_N  \int
\frac{d\Omega_N}{16\pi^2}\frac{|{\bf p}_N|}{m_{D\ell}}
 F_{BD}^+  F_{DD}^+
\label{Mabs}\\
   && \bar u_{\bar\ell}(l_2) (\not\hspace{-2pt}p_D +
\not\hspace{-2pt}p_{D^0} ) (1+\gamma_5)(\not\hspace{-2pt}p_N +
m_N) (\not\hspace{-2pt}p_{D^0} + \not\hspace{-2pt}p_B )
(1-\gamma_5) v_\ell (l_1),\nonumber
\end{eqnarray}
This angular integral is quite simple, because in the $D$-$N$
frame the energy of every particle in the process is fixed. The
subsequent steps to obtain the decay rate are straightforward and
described in the Appendix. The expression for the rate is thus the
integral [see Eq.~\eqref{Rate_light}]:

\begin{equation}
\Gamma (B^+ \to D^- \ell^+ \ell^-)
  =
  \frac{G_F^4}{(16\pi^2)^2} |V_{cb}V_{ud}|^2  F_{BD}^{+ 2}
F_{DD}^{+ 2} \frac{|U_{N\ell}^2 m_N|^2}{m_B ^2}
 \int\limits_{(m_D + m_\ell)}\limits^{(m_B-m_\ell)}  \frac{d m_{D\ell} }{2\pi}  \
 \frac{|{\bf p}_N|^2}{m_{D\ell}^2} |{\bf \tilde l}_1|\  |{\bf  l}_2|
 \  \times {\cal R},
 \label{Rate_light1}
\end{equation}
where $|{\bf p}_N|$, $|{\bf \tilde l}_1|$ and $|{\bf  l}_2|$ are
the 3-momenta of the neutrino and leptons (given in the Appendix)
and ${\cal R}$ is a quantity of dimension $m^6$ shown in
Eq.~\eqref{Rvalue}. The integral can be easily done numerically,
which we do considering a $D$ meson in the intermediate and final
states ($b\to c$ transition), or alternatively a pion ($b\to u$
transition).

Notice that by assuming the form factors to be constant unity we
are overestimating the process, while by neglecting the $F^-$ form
factors and the dispersive part of the amplitude we may be
inducing an uncertainty of an order of magnitude. Within our
approximations, in both cases the results for the branching ratios
are extremely small:
\begin{eqnarray}
Br(B^+\to D^- \ell^+ \ell^+) &\sim& 1.2\times 10^{-31}
 \left(\frac{U_{N\ell}^2 m_N}{1~\textrm{eV}}\right)^2
 ,\label{BDee}
 \\
Br(B^+\to \pi^-\ell^+ \ell^+) &\sim& 2.3\times 10^{-33}
\left(\frac{U_{N\ell}^2 m_N}{1~\textrm{eV}}\right)^2 ,\label{BPee}
\end{eqnarray}
where we used the values of the CKM elements shown in Table~\ref{table1},
and also $\Gamma_B = 4.0\times 10^{-13}$ GeV.
We can compare these results with those of A.~Ali {\it et al.} \cite{Ali}, who
considered the $s$-type diagram only. In our notation, their result for
$Br(B^+\to\pi^-e^+e^+)$ becomes $(0.3-1.8)\times 10^{-35}({U_{N\ell}^2
m_N}/eV)^2$, which is two orders of magnitude smaller than Eq.~\eqref{BPee}.

Nevertheless, we expect our results to be just rough
estimates within one or two orders of magnitude, as we have taken
the form factors $F^+_{BD} \sim F^+_{DD}$ to be unity, and we have
neglected the form factors $F^-_{BD}$ and $F^-_{DD}$ altogether.
In general,
the form factors $F^+$ are expected to be unity at most at the kinematical end point
where the two meson wave functions could overlap completely (provided they have the same shape),
but it should be smaller for all other $q^2$ values.

Taking for $F^+$
an average value of {e.g.} $0.3$ instead of unity, our calculated rates get reduced  by a factor
$(F^+)^4\sim 10^{-2}$, reducing Eq.~\eqref{BPee} to a value comparable with the result of Ali {\it et al.}
%
%Indeed,  Eq.~\eqref{BPee} can be compared with the result of
%A.~Ali {\it et al.} \cite{Ali}, which

%The large theoretical uncertainty of our estimate can only be reduced by resorting to models of
%the form factors, and so it cannot clearly resolve the issue of dominance of either diagram in Fig.~\ref{fig1} for the case of light neutrinos,

Accordingly, in the case of light neutrinos, our crude estimate cannot clearly show the dominance of the $t$-type diagram. However, it does show at least that a calculation based purely on the $s$-type diagram may be an underestimation \cite{Controversial}. It also shows that this potential underestimation is hardly more than two orders of magnitude, keeping these branching ratios still beyond the reach of foreseen experiments, as concluded in Ref.~\cite{Ali}.

%  END OF THE MODIFICATION TO COMPARE WITH ALI.

To estimate the actual range of these branching ratios we would
need to have estimates of the neutrino masses and mixings as well.
Using the standard parametrization of the PMNS neutrino
mixing matrix multiplied by a $3\times 3$ Majorana phase matrix,
the term $U_{N\ell}^2 m_N$ can be explicitly written in terms of
three light neutrino masses, three neutrino mixing angles, two
Majorana phases and one Dirac phase.
Since the sign of $\Delta m_{31}^2$ is not determined from the
existing data, there are two possible neutrino mass hierarchies,
one called normal $(m_3 > m_{1,2})$ and the other inverted $(m_3 <
m_{1,2})$. The size of the term $U_{N\ell}^2 m_N$ in general
depends on the mass hierarchy.
If we consider standard neutrinos, we know that $m_\nu < 2$ eV,
and we can roughly use $U_{\nu \ell}\sim {\cal O}(1)$ for either
$\ell = e$ or $\mu$, in consistency with oscillation experiments.
We then get branching ratios smaller than $10^{-31}$ and
$10^{-33}$, respectively, values which are prohibitively small for
any foreseen experiment.
On the other hand, if we consider heavier neutrinos (but still
lighter than $m_\pi$), $i.e.$ $m_N\sim 100$ MeV, the results could
be more promising, but in those cases we should use the mixings of
standard with extra neutrinos, which are suppressed: $U_{N e}^2,
U_{N\mu}^2< 0.002$, \cite{Nardi} so the resulting branching ratios
have the upper bounds $10^{-21}$ and $10^{-23}$, respectively,
which are still prohibitively small.

As a final remark, we want to comment on the assumptions involved
in this calculation. First, the fundamental process at the quark
level (see Fig.~1) with two electroweak vertices has been modeled
as a process with hadrons and leptons, where a single long
distance contribution (an intermediate state with a meson and a
neutrino on shell) is supposed to dominate; we have thus neglected
other possible intermediate hadronic states (e.g.  excitations of
the intermediate meson and multimeson states) as well as a short
distance contribution where both weak vertices coalesce into a
single one \cite{double_beta}.
We have assumed the dominance of the single $D$-$N$ intermediate
channel as it goes on its mass shell. Another assumption was to
consider the absorptive part as representative of the full
amplitude; since we are only after an order of magnitude estimate,
this is likely to be a good assumption, again due to the resonant
character of the intermediate state as it goes on its mass shell.
Within the hadronic approximation for the weak currents,
we took into account just one of the form factors of each hadronic
current, and assumed it to be constant (unity) within the whole
dynamic range. In principle one can expect the form factor to be
unity at most, as explained before;
taking the $q^2$ dependence into account one should then obtain a
lower value for the rate, but as we have seen, it is unlikely for this effect to change the result
by more than two orders of magnitude. These approximations are
therefore consistent with the level of precision we seek.

%%%%%%%%%%%%%%%%%%%%%%%%%%%%%%%%%%%%%
%
%                         SUBSECTION 2.2    INTERMEDIATE CASE
%
%%%%%%%%%%%%%%%%%%%%%%%%%%%%%%%%%%%%%

\subsection{ The case of intermediate mass neutrinos ($ m_\pi  <m_N < m_{B_c}$)}

\begin{figure}[t]
  \centering
  \includegraphics[scale=0.8]{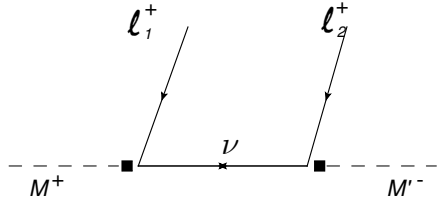}
  \caption{The dominating diagram  (plus diagram with leptons
  exchanged if they are identical)
   in an effective meson theory for$M^+\to {M'}^-\ell^+\ell^+$, mediated by Majorana
  neutrinos with mass in the range between $m_{M'}$ and $m_M$.}
  \label{fig3}
\end{figure}

In contrast to the previous case, the process $M^+\to {M'}^-
\ell^+\ell^+$ in the case of Majorana neutrinos with masses in the
intermediate range $m_{M'}< m_N < m_M$ is dominated by the $s$-type amplitude
of Fig.~\ref{fig1}, corresponding at the meson level to Fig.~\ref{fig3}, as the neutrino in the
intermediate $s$-channel goes into its mass shell.
As stated in the Introduction, Majorana neutrinos with such masses
must be sterile and should originate from new physics beyond the SM.

Since there are two identical leptons in the final state, one must
also consider the diagram with crossed leptons and then integrate
over half the phase space.
However, for the case of the intermediate neutrino on mass shell,
the result is the same as using a single diagram, as if the
leptons were distinct, as shown in Fig.~\ref{fig3}.
The  effective amplitude then is:
\begin{equation}
{\cal M}= \frac{G_F^2}{2} U_{N \ell}^{\ast 2}\
{V_{qQ}^\ast}{V_{q_2 q_1}^\ast}
 f_M f_{M'} \ \frac{\tilde M}{(p_N^2-m_N^2)+i m_N \Gamma_N}
 \label{amplitude1}
\end{equation}
where $U_{N \ell}$ and $V_{q_i q_j}$  are  the PMNS lepton mixing
and CKM quark mixing elements, respectively, $f_M$, $f_{M'}$ are
the meson decay constants, and we define $\tilde M$ as the reduced
matrix element that contains all the spinor structure of the
amplitude:
\begin{equation}
\tilde{\cal M} =\lambda_N \  \bar u_{\bar \ell}(l_1)\,\pslash_M
(1+\gamma_5)\, (\pslash_N +m_N)  \, \pslash_{M'} (1-\gamma_5)
v(l_2) \label{amplitude2}
\end{equation}
where the notation is the same as in Eq.~\eqref{Mabs}.

The decay rate we seek is then given by $\Gamma(M^+\to
{M'}^-\ell^+\ell^+) = (1/2m_M)\int d{\rm ps}_3 |{\cal M}|^2$,
where $d{\rm ps}_3$ is the final 3-particle phase space. The
calculation of the squared matrix element and the integration over
the final phase space are shown in Appendix 2, resulting in the
following expression [see Eq.~\eqref{Interm}]:
\begin{eqnarray}
\Gamma(M\to M' \ell^+ \ell^+)
 &=&
\frac{G_F^4}{32\pi^2 m_M}  f_M^2 f_{M'}^2 |{V_{qQ}}{V_{q_2
q_1}}|^2  \frac{|{U_{N \ell}}|^4}{m_N \Gamma_N}
  \frac{|\tilde{\bf l}_1|}{m_M}
  \frac{|{\bf l}_2|}{m_N}
\label{rate} \\ && \times
\left\{
 (m_N^2 + m_\ell^2)m_M^2 - (m_N^2 - m_\ell^2)^2 \right\}
 \left\{
 (m_N^2 - m_\ell^2)^2 - (m_N^2 + m_\ell^2) m_{M'}^2 \right\},
\nonumber
\end{eqnarray}
where $|\tilde{\bf l}_1|$ and $|{\bf l}_2|$ are the 3-momenta of
the first electron in the $M$ meson rest frame and of the second
electron in the neutrino rest frame, respectively.

Before we can use this expression, we also need a theoretical
expression for $\Gamma_N$, the total decay width of the
intermediate Majorana neutrino, in terms of the same neutrino
parameters we have just used.
The total width $\Gamma_N$ can be estimated by comparing the decay
modes of $N$ with those of the $\tau^-$ lepton, where
$\Gamma_{\tau} \propto m_{\tau}^5$. Both $N$ and $\tau^-$ decay
via the same type of diagrams and couplings, but there are a few
differences: (a) $N$ has a different mass (thus $\Gamma_N \propto
m_N^5$); (b) $\Gamma_N$ has an additional factor of two due to the
Majorana character of $N$ (unlike $\tau^-$ which is a Dirac
particle), because it decays with equal probability into both
$(\ell^{\prime -} + {\rm rest}^+)$ and $(\ell^{\prime +} +
{\overline {\rm rest}}^-)$; (c) $\Gamma_N$ has an additional
mixing factor $|U_{N \ell^{\prime}}|^2$. Therefore:
\begin{equation}
\Gamma_N \approx  2 \sum_{\ell'} |U_{N \ell'}|^2
\left(\frac{m_N}{m_\tau}\right)^5 \times \Gamma_\tau,
\label{GammaNtot}
\end{equation}
This expression for $\Gamma_N$ is a good approximation when $m_N$
is near $2$ GeV; in this case the decay channels of $N$ are those
of $\tau$, where the virtual $W$ boson produces $e^- {\overline
\nu}_e, \mu^- {\overline\nu}_{\mu}$ and $d {\overline u}$ (the
last channel is actually a set of three, due to color). However,
for $m_N
> 2$ GeV, the additional channels $\tau^- {\overline\nu}_{\tau}$ and
$s  {\overline c}$ open, increasing the expression in
Eq.~(\ref{GammaNtot}) by up to a factor $\approx 1.5$, including
phase space suppression due to the masses of the products.
Consequently, using Eq.~(\ref{GammaNtot}) in Eq.~(\ref{rate}) may
oversetimate the rates by at most $\sim 30 \%$. We will thus use
Eq.~(\ref{GammaNtot}) in the estimation of the LNV rates, but
keeping in mind that a correction in $\Gamma_N$ should be included
in a more refined study.

Accordingly, and if we neglect the charged lepton mass,
Eq.~\eqref{rate} turns into:
%
%
%For neutrino masses above 1 GeV, the neutrino should have decay
%channels similar to those of a tau lepton, and so the expression
%for the width should be also similar, except for $m_N \neq m_\tau$
%and the flavor mixing suppression factor
% $|U_{N \ell}|^2$. Consequently, since $\Gamma_\tau \sim m_\tau^5$, then
%\[
%\Gamma_N \sim  \sum_{\ell'} |U_{N \ell'}|^2
%\left(\frac{m_N}{m_\tau}\right)^5 \times \Gamma_\tau,
%\]
%where the sum is over all standard charged leptons that are
%kinematically allowed in the final state.
%
%
\begin{equation}
\Gamma(M\to M' \ell^+ \ell^+) \approx \frac{1}{128\pi^2} {G_F^4}
f_M^2 f_{M'}^2 |{V_{qQ}}{V_{q_2 q_1}}|^2  \frac{ |U_{N
\ell}|^4}{\sum_{\ell'} |U_{N \ell'}|^2}
 \frac{m_M m_\tau^5 }{2\Gamma_\tau}\left(1-\frac{m_{M'}^2}{m_N^2}\right)^2
 \left(1-\frac{m_N^2}{m_M^2}\right)^2 .
 \label{rate2}
\end{equation}
Here we will use $m_\tau = 1.77$ GeV and $\Gamma_\tau = 2.3\cdot
10^{-12}$ GeV \cite{PDG2008}. Eq.~\eqref{rate2} is valid for $m_N$ in the range
$m_{M'}~<~m_N~<~m_M$, it vanishes at the two endpoints of this
range, and reaches its maximum at $m_N=\sqrt{m_M\cdot m_{M'}}$,
where $(1-m_{M'}^2/m_N^2)^2(1-m_N^2/m_M^2)^2\to(1-m_{M'}/m_M)^4$.

Consequently, these suppressed non-standard decays can impose more
or less stringent bounds on the mixing elements between the standard
leptons and extra neutrinos, $|U_{N\ell}|$, depending on the
Majorana neutrino mass. In particular, the
non-observation of these processes defines $m_N$-dependent upper
bounds for the corresponding $|U_{N\ell}|$.

\begin{figure}[htb]
\centering
  \includegraphics[width=8.cm]{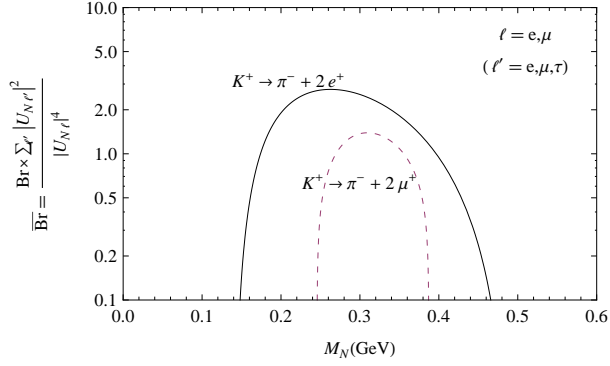}
\vspace{-0.4cm} \caption{Branching ratios for $K^+ \to \pi^-
\ell^+ \ell^+$ ($\ell = e, \mu$) as functions of the exchanged
neutrino mass $m_N$ in the range $m_\pi<m_N<m_K$, with the lepton
mixing factor, ${ |U_{N e}|^4}/{\sum_{\ell'} |U_{N \ell'}|^2}$, divided out.} \label{plK}
\end{figure}

\begin{figure}[t]
\begin{minipage}[b]{.49\linewidth}
\centering
 \includegraphics[width=\linewidth]{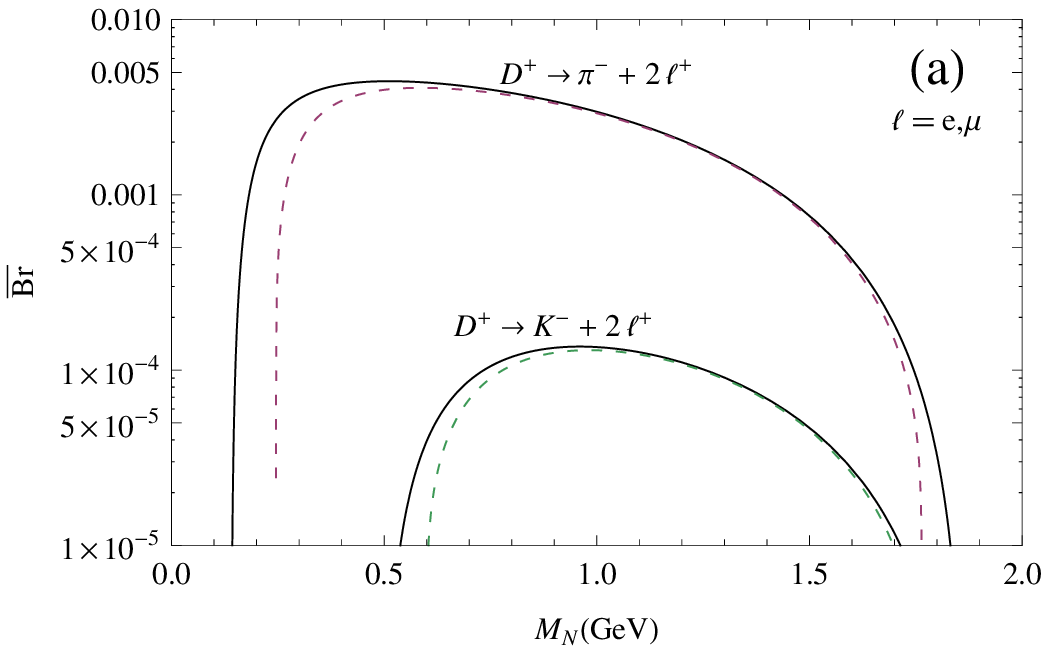}
\end{minipage}
\begin{minipage}[b]{.49\linewidth}
\centering
 \includegraphics[width=\linewidth]{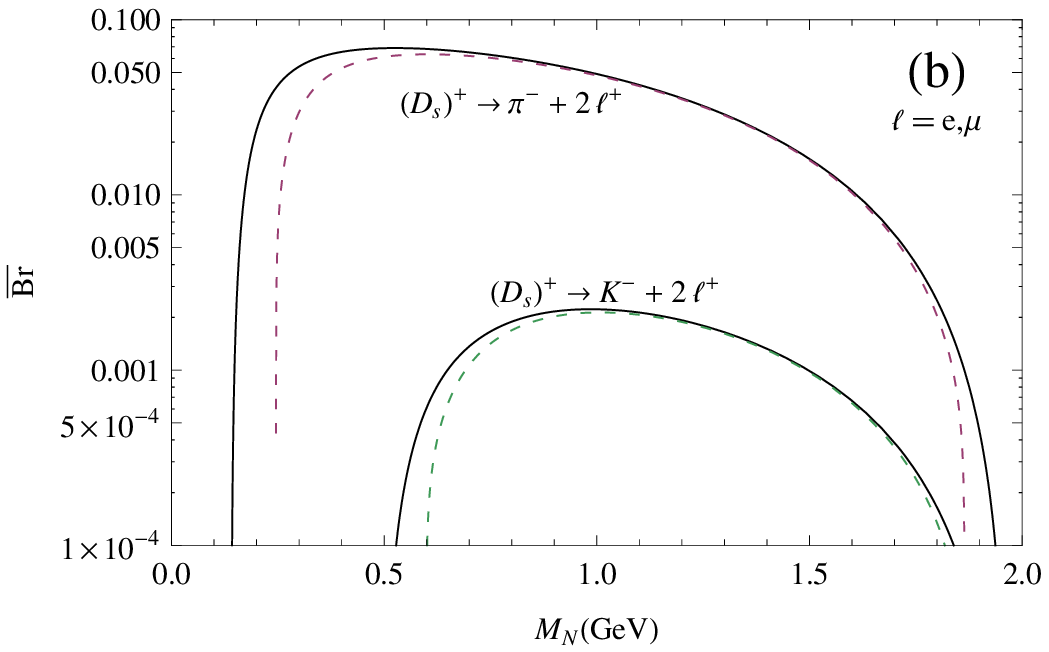}
\end{minipage}
\vspace{-0.2cm} \caption{Branching ratios for (a) $D^+$ decays and
(b) $D_s^+$ decays, as functions of the neutrino mass $m_N$, with
the lepton mixing factor divided out as in Fig.~\ref{plK}. The
full lines correspond to $\ell = e$ and the dashed lines to $\ell
= \mu$.} \label{plDplDs}
\end{figure}

\begin{figure}[ht]
\begin{minipage}[b]{.49\linewidth}
\centering
 \includegraphics[width=\linewidth]{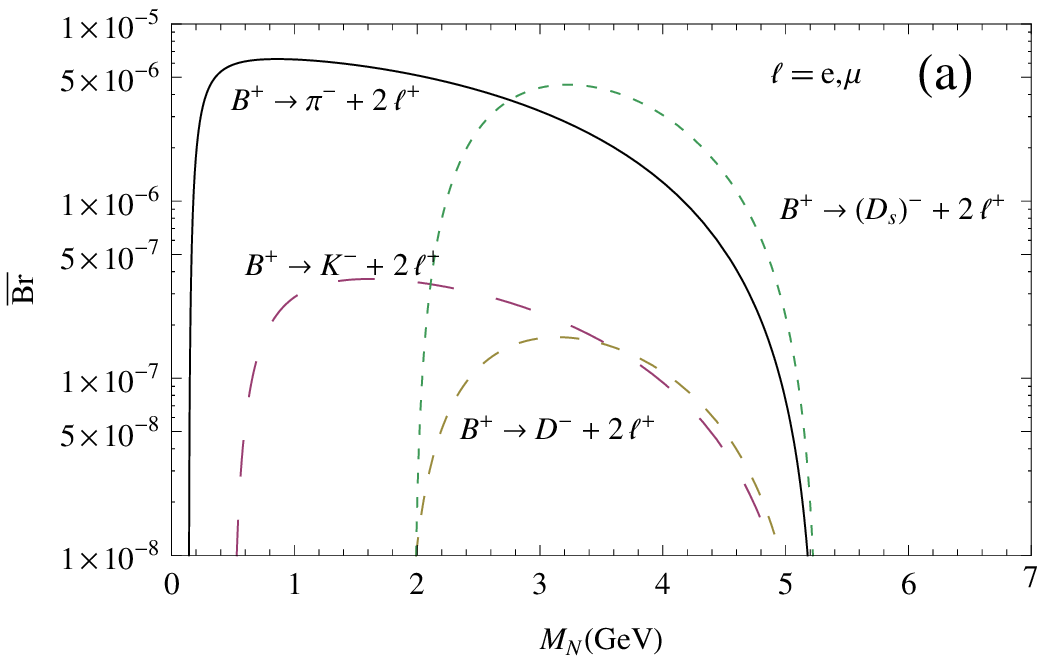}
\end{minipage}
\begin{minipage}[b]{.49\linewidth}
\centering
 \includegraphics[width=\linewidth]{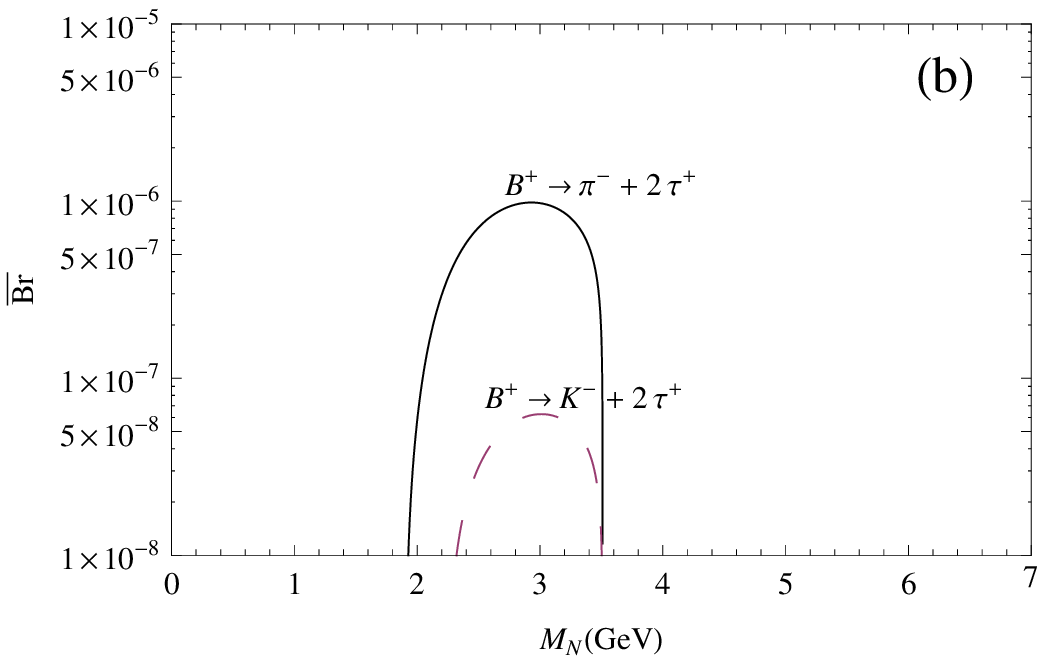}
\end{minipage}
\vspace{-0.2cm} \caption{Branching ratios for $B^+ \to M^{\prime
-} \ell^+ \ell^+$ as functions of the neutrino mass $m_N$, with
the lepton mixing factor divided out as in Fig.~\ref{plK}. The
produced pseudoscalars are $M^{\prime} = \pi, K, D, D_s$. (a) The
case of leptons with negligible mass ($\ell = e, \mu$); (b) the case $\ell =
\tau$ (here $M^{\prime} = D, D_s$ are kinematically forbidden).}
\label{plB}
\end{figure}

\begin{figure}[ht]
\begin{minipage}[b]{.49\linewidth}
\centering
 \includegraphics[width=\linewidth]{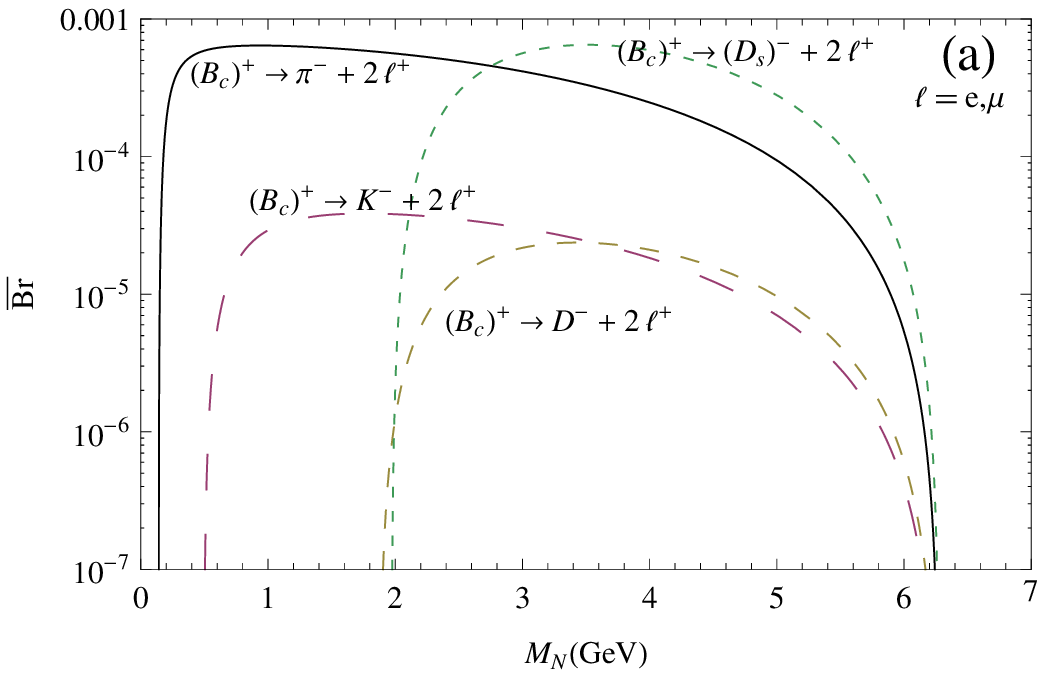}
\end{minipage}
\begin{minipage}[b]{.49\linewidth}
\centering
 \includegraphics[width=\linewidth]{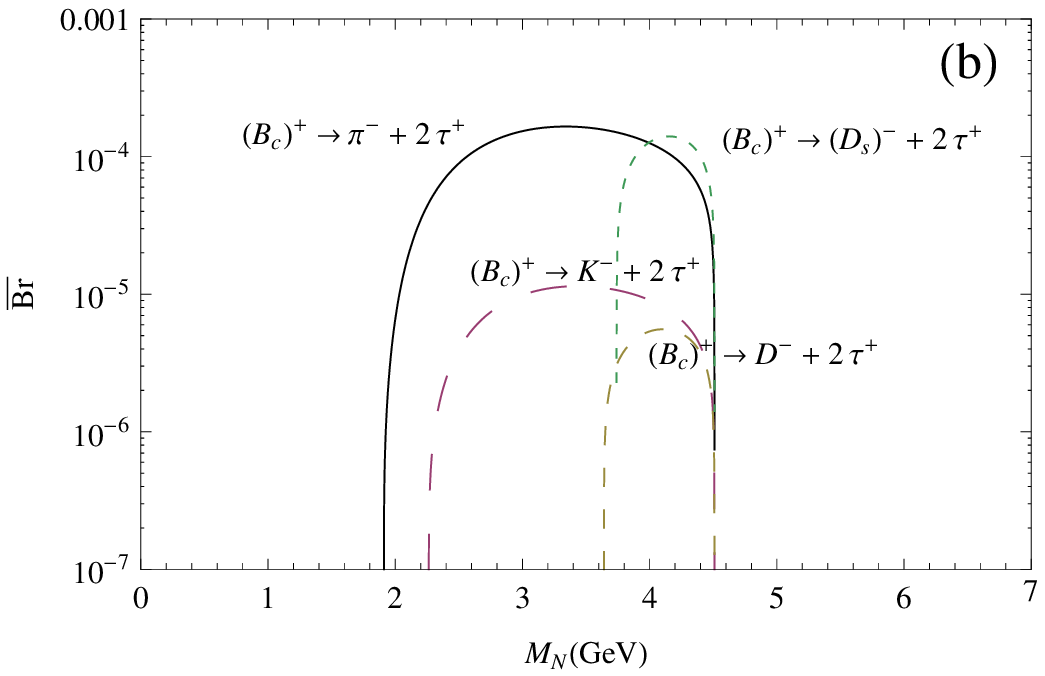}
\end{minipage}
\vspace{-0.2cm} \caption{Branching ratios for $B_c \to M^{\prime
-} \ell^+ \ell^+$ as functions of the neutrino mass $m_N$, with
the lepton mixing factor divided out as in Fig.~\ref{plK}. The
produced pseudoscalars are $M^{\prime} = \pi, K, D, D_s$. (a) The
case of leptons with negligible mass ($\ell = e, \mu$); (b) the case $\ell =
\tau$.} \label{plBc}
\end{figure}

In Figs.~\ref{plK}--\ref{plBc} we show the branching ratios for
the decays $K^+\to \pi^- \ell^+ \ell^+$, $D^+\to M^{\prime -}
\ell^+ \ell^+$, $D^+_s\to M^{\prime -} \ell^+ \ell^+$, $B^+\to D^-
\ell^+ \ell^+$ and $B^+_c\to M^{\prime -} \ell^+ \ell^+$ as
functions of $m_N$, where the bounds on the mixings $|U_{N\ell}|$
can be deduced also as functions of $m_N$.

\begin{table}[htb]
\caption{The coefficients ${\cal C}$ appearing in
Eq.~\eqref{rate4} for the maximal branching ratio,  and the
neutrino mass $m_N$ at which the maximum is reached, for various
decays $M^+ \to M^{\prime -} \ell^+ \ell^+$, where $m_\ell$ can be
neglected. In the last column, the expected upper bound on the
branching ratios, provided $|U_{N\ell}|^2\sim 10^{-6}$ or
$10^{-7}$, for $m_N\sim 0.1$ GeV or $\sim 1$ GeV, respectively.}
\label{tabcalC}
%\begin{ruledtabular}
\begin{tabular*}{0.70\textwidth}{@{\extracolsep{\fill}} |llll| }
\hline decay & ${\cal C}$  &
 $m_N$ at maximum & $Br<$
\\
\hline\hline
 $K^+ \to \pi^- \ell^+ \ell^+$ & $2.8$  & 0.26~GeV
 & $2.8 \cdot 10^{-6}$
\\
\hline
 $D^+ \to \pi^- \ell^+ \ell^+$ & $4.5 \cdot 10^{-3}$  & 0.51~GeV
 & $4.5 \cdot 10^{-10}$
\\
 $D^+\to K^-\ell^+\ell^+$ & $1.4\cdot 10^{-4}$ &  0.96~GeV
 & $1.4 \cdot 10^{-11}$
\\
\hline
 $D_s^+ \to \pi^- \ell^+ \ell^+$ & $6.9 \cdot 10^{-2}$  & 0.53~GeV
 & $6.9 \cdot 10^{-9}$
\\
 $D_s^+ \to K^- \ell^+ \ell^+$ & $2.2 \cdot 10^{-3}$ & 0.99~GeV
 & $2.2 \cdot 10^{-10}$
\\
 $D_s^+ \to D^- \ell^+ \ell^+$ & $8.5\cdot 10^{-8}$ & 1.92~GeV
 & $8.5 \cdot 10^{-15}$
\\
\hline
 $B^+ \to \pi^-\ell^+ \ell^+$ & $6.3 \cdot 10^{-6}$  & 0.86~GeV
 & $6.3\cdot 10^{-13}$
\\
 $B^+ \to K^- \ell^+ \ell^+$ & $3.6 \cdot 10^{-7}$ &  1.61~GeV
 & $3.6\cdot 10^{-14}$
\\
 $B^+\to D^-\ell^+\ell^+$ & $1.7\cdot 10^{-7}$ & 3.14~GeV
 & $1.7\cdot 10^{-14}$
\\
 $B^+\to D_s^-\ell^+\ell^+$ & $4.5\cdot 10^{-6}$ & 3.23~GeV
 & $4.5 \cdot 10^{-13}$
\\
\hline
 $B_c^+\to\pi^- \ell^+\ell^+$ & $6.4\cdot 10^{-4}$  & 0.94~GeV
 & $6.4\cdot 10^{-11}$
\\
 $B_c^+ \to K^-\ell^+\ell^+$ & $3.9 \cdot 10^{-5}$ & 1.76~GeV
 & $3.9\cdot 10^{-12}$
\\
 $B_c^+ \to D^- \ell^+ \ell^+$ &  $2.4 \cdot 10^{-5}$ & 3.43~GeV
 & $ 2.4\cdot 10^{-12}$
\\
 $B_c^+ \to D_s^- \ell^+ \ell^+$ &  $6.5\cdot 10^{-4}$ & 3.52~GeV
 & $6.5\cdot 10^{-11}$
\\
 $B_c^+ \to B^- \ell^+ \ell^+$ &  $1.6\cdot 10^{-11}$ & 5.76~GeV
 & $1.6 \cdot 10^{-18}$
\\
\hline
\end{tabular*}
%\end{ruledtabular}
\end{table}

Let us consider the decay $B^+\to D^- e^+ e^+$ as an example. Here
we must use $V_{qQ}\to V_{ub}$ and $V_{q_1 q_2}\to
V_{cd}$ as inputs, as well as $f_B$ and $f_D$ (see Table~\ref{table1}) and $\Gamma_B =
4.0\cdot 10^{-13}$ GeV \cite{PDG2008}.
The branching ratio for this process as a function of $m_N$ is shown
in Fig.~\ref{plB}(a), lower dashed line, and reaches a maximum:
\begin{equation}
{\rm Br}_{\rm max} (B^+\to D^- e^+ e^+) = 3\cdot 10 ^{-7} \times
\frac{ |U_{N e}|^4}{\sum_{\ell'} |U_{N \ell'}|^2} \quad {\rm at}\
\  m_N \sim 3~{\rm GeV}. \label{ratemax1}
\end{equation}
%
%If $|U_{N e}|$ happens to be the largest among the $|U_{N\ell'}|$
%($\ell'= e, \mu, \tau$), then the factor $|U_{N e}|^4
%/(\sum_{\ell'} |U_{N \ell'}|^2)$ reduces to $\approx |U_{N e}|^2$.
%
This expression just gives the maximal possible value of this
branching ratio, which occurs only if $m_N$ happens to be near 3
GeV, but for other values of $m_N$, it could be much smaller, as shown in
Fig.~\ref{plB}(a).

Analogous to Eq.~\eqref{ratemax1},  the maximal
branching ratio of any of the other decays has the form:
\begin{equation}
{\rm Br}_{\rm max} (M^+\to {M'}^-  \ell^+ \ell^+) =
 {\cal C}
\times \frac{ |U_{N \ell}|^4}{\sum_{\ell'} |U_{N \ell'}|^2} .
\label{rate4}
\end{equation}
Table \ref{tabcalC} shows the coefficient ${\cal C}$ appearing in
Eq.~\eqref{rate4}, for the different branching ratios, and the
value of the corresponding neutrino mass $m_N$ at which the
maximal branching ratio  is reached.

Accordingly, an experimental upper bound on the branching ratio
for $M^+\to {M'}^- \ell^+ \ell^+$ imposes an upper bound on the
leptonic mixings $|U_{N\ell}|$,  bound that strongly depends on
the neutrino mass $m_N$, and which is most stringent if $m_N \sim
\sqrt{m_M\cdot m_{M'}}$, where the branching ratio is maximal. For
$m_N$ away from that value, the upper bounds imposed on the
mixings become much less stringent.

{}From the ${\cal C}$ values in Table~\ref{tabcalC} one can read the potential
of different processes to set upper bounds on the lepton mixing
elements $|U_{N\ell}|$, for different neutrino masses $m_N$. For a
given experimental upper bound of a branching ratio, the larger
the ${\cal C}$ coefficient, the more stringent the upper bound
that can be imposed on $|U_{N\ell}|$, provided the neutrino mass
is near the indicated value where the theoretical branching ratio
is maximal.

{}From Eq.~(\ref{rate4}) it is clear that the bounds on the
mixings imposed from these decays appear in the combination
\begin{equation}
\frac{ |U_{N \ell}|^4 }{|U_{N e}|^2 + |U_{N \mu}|^2 + |U_{N
\tau}|^2},\quad \ell= e,\mu \textrm { or } \tau,
\label{mixfactor}
\end{equation}
not just $|U_{N \ell}|$. Only if $|U_{N \ell}|$ is much larger
than the other mixings, then this expression reduces to $|U_{N
\ell}|^2$. Otherwise, one must use the bounds on $Br(M\to {M'}^-
\ell^+\ell^+)$ for a given meson pair $M$ and $M'$, but for
\emph{all} lepton flavors $\ell= e,\mu, \tau$, in order to
disentangle the bounds for each of the mixings $|U_{N \ell}|$.
Moreover, these bounds will depend on $m_N$, since the relation
between the branching ratios and the mixings depend on $m_N$, as
it was already mentioned and shown in Figs.~\ref{plK}--\ref{plBc}.

On the other hand,  to explore the prospects of experimentally
observing any of these processes, one needs at least an estimate
of the $|U_{N\ell}|$ elements. Present upper bounds on the
heavy-to-light neutrino mixing $| U_{N \ell} |^2$ for
$\ell=e,\mu$, vary considerably with the neutrino mass, but are
typically in the range $| U_{N \ell} |^2< 10^{-4},\ 10^{-6}, \
10^{-7}$, for $m_N\sim 10$ MeV, $100$ MeV, $1$ GeV, respectively
(pp.~546-548 in Ref.~\cite{PDG2008}).
We have then listed in the last column of Table~\ref{tabcalC} the expected
upper bound on the branching ratios, provided the mixing elements
have the values just mentioned.

%%%%%%%%%%%%%%%%%%%%%%%%%%%%%%%%%%%%%
%
%                         SUBSECTION 2.3    HEAVY CASE
%
%%%%%%%%%%%%%%%%%%%%%%%%%%%%%%%%%%%%%

\subsection{ The case of heavy neutrinos ($m_N > m_{B_c}$)}

\begin{figure}[t]
  \centering
  \includegraphics[scale=0.8]{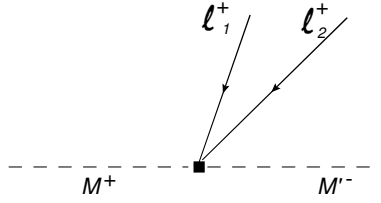}
\caption{The diagram in an effective meson theory for $M^+\to
M^{\prime -} e^+e^+$, when the neutrino mass is much larger than
that of the decaying meson, which is resulted from  the four
amplitudes of Fig. 1.}
  \label{fig4}
\end{figure}

If neutrinos happen to be much heavier than the decaying meson,
then in general both diagrams in Fig.~1 contribute with more or
less the same strength, and reduce at the meson level to a single
point-like interaction diagram as shown in Fig.~\ref{fig4}.
The vertex in Fig.~\ref{fig4} represents the double weak
interaction shown in Fig.~1, where the neutrino line as well as
all other internal lines have been reduced to a point. At the
meson level, the specific tensor structure of this four-particle
vertex cannot be selected among all the general possibilities, so
we start from the fundamental quark and lepton interactions as
shown in Fig.~1 and exhibit the approximations involved to get to
the leading term at the meson level. These details are presented
in Appendix C. Our model of the dynamics in this case
is equivalent to that of Ali {\it et al.} \cite{Ali}.
In summary, if we can approximate the hadronic
tensor by the product of two currents, factorized by a vacuum
insertion, the squared amplitude is then given in terms of the mesons'
decay constants and the kinematics of mesons and leptons separate
into independent factors (see Eq.~\ref{M2heavy}):
\[
|{\cal M}|^2 \sim f_M^2 f_{M'}^2 (p_M\cdot p_{M'})^2  (\ell_1 \cdot \ell_2).
\]
The decay rate then becomes (see Eq.~\ref{G2}):
\begin{eqnarray}
 \Gamma(M^+ &&\to M^{\prime -} \ell^+ \ell^+)
 =
  \frac{G_F^4}{128\pi^3 }
 \left|\frac{U_{N \ell}^{*2}}{m_N}\right|^2
\left| V_{q Q}^* V_{q_1 q_2}^* + \frac{V_{q_1 Q}^* V_{q
q_2}^*}{N_c}  \right|^2  f_M^2 f_{M'}^2\, m_M^3
 \label{G22} \\
&& \int \limits_{4 m_\ell^2}\limits^{(m_M-m_{M'})^2}
 d m_{\ell\ell}^2 \,
\lambda^{1/2}( 1,\frac{m_{M'}^2}{m_M^2}, \frac{m_{\ell\ell}^2}{m_M^2}) \,
\lambda^{1/2}(1,\frac{m_\ell^2}{m_{\ell\ell}^2}, \frac{m_{\ell}^2}{m_{\ell\ell}^2})\,
 \left(1+\frac{m_{M'}^2}{m_M^2}-\frac{m_{\ell\ell}^2}{m_M^2}\right)^2\,
(m_{\ell\ell}^2 - 2 m_\ell^2), \nonumber
\end{eqnarray}
where the function $\lambda(x,y,z)$ is defined in
Eq.~\eqref{lambda1}. This expression exactly coincides with the expression
obtained in Ref.~\cite{Ali} for the heavy neutrino cases.

The integral above can be easily done
numerically. In order to do the phenomenology, we set a fiducial
value for the neutrino mass $m_N = 100$ GeV, and a corresponding
mixing element $|U_{N\ell}|^2 = 10^{-2}$ and express the branching
fraction of this decay in terms of a dimensionless quantity ${\cal
B}$, whose value, according to Eq.~\eqref{G22}, is determined by
the masses of the external particles:
\begin{equation}
{\rm Br} (M^+ \to M^{\prime -} \ell^+ \ell^+) \equiv
\frac{ \Gamma(M^+ \to M^{\prime -} \ell^+ \ell^+)}{\Gamma_M} =
{\cal B} \times \left( \frac{100 \ {\rm GeV}}{m_{N}} \right)^2
\left(\frac{|U_{N \ell}|^2}{10^{-2}} \right)^2 . \label{resBr1}
\end{equation}

For the case $B^+\to D^-\ell^+\ell^+$ we must use $f_M = f_{B^+}$ and
$f_{M^{\prime}} = f_{D^-}$. Using the values shown in Table~\ref{table1}, as well as
$\Gamma_{B^+} = 4.0 \cdot 10^{-13}$ \cite{PDG2008}, the result is:
\begin{eqnarray} {\rm Br}(B^+ \to D^- \ell^+ \ell^+) & \approx &
1.1 \cdot 10^{-22} \times \left( \frac{100 \ {\rm GeV}}{m_{N}}
\right)^2 \left(\frac{|U_{N \ell}|^2}{10^{-2}} \right)^2 \ .
\label{resBBr}
\end{eqnarray}
Similar results can be obtained for decays of other mesons $M^+ =
K^+, D^+, D_s^+, B_c^+$. The coefficients ${\cal
B}$ for various decays are given in Table \ref{tabcalA}.
\begin{table}
\caption{Branching ratio coefficients ${\cal B}$ appearing in
Eq.~(\ref{resBr1}), for various decays $M^+ \to M^{\prime -}
\ell^+ \ell^+$, if the process is dominated by heavy neutrinos
($m_N \gg m_M$). ${\cal B}$ values correspond to branching ratios if $m_N = 100$ GeV and
$|U_{N\ell}|^2 = 10^{-2}$.
All three lepton flavors are considered ($\ell = e, \mu,  \tau$).
Entries are empty for decays that are kinematically forbidden.} \label{tabcalA}
%\begin{ruledtabular}
\begin{tabular*}{0.70\textwidth}{@{\extracolsep{\fill}} |llll| }
\hline
 decay & ${\cal B} (\ell=e)$  & ${\cal B} (\ell=\mu)$ &
 ${\cal B} (\ell=\tau)$
\\
\hline\hline
 $K^+ \to \pi^- \ell^+ \ell^+$ & $8.47 \cdot 10^{-24}$ &
$2.44 \cdot 10^{-24}$ & -
\\
\hline $D^+ \to \pi^- \ell^+ \ell^+$ & $1.90 \cdot 10^{-23}$
& $1.78 \cdot 10^{-23}$  & -
\\
$D^+ \to K^- \ell^+ \ell^+$ & $1.58 \cdot 10^{-23}$
& $1.47 \cdot 10^{-23}$ & -
\\
\hline $D_s^+ \to \pi^- \ell^+ \ell^+$ & $2.14 \cdot 10^{-22}$
& $2.02 \cdot 10^{-22}$ & -
\\
$D_s^+ \to K^- \ell^+ \ell^+$ & $2.46 \cdot 10^{-23}$
& $2.30 \cdot 10^{-23}$ & -
\\
$D_s^+ \to D^- \ell^+ \ell^+$ & $6.99 \cdot 10^{-28}$
& - & -
\\
\hline $B^+ \to \pi^- \ell^+ \ell^+$ & $1.13 \cdot 10^{-23}$
& $1.12 \cdot 10^{-23}$ &  $7.42 \cdot 10^{-25}$
\\
$B^+ \to K^- \ell^+ \ell^+$ & $8.44 \cdot 10^{-25}$
& $8.37 \cdot 10^{-25}$ &  $5.01 \cdot 10^{-26}$
\\
$B^+ \to D^- \ell^+ \ell^+$ & $1.02 \cdot 10^{-22}$
 & $1.01 \cdot 10^{-22}$ & -
\\
$B^+ \to D_s^- \ell^+ \ell^+$ &  $5.02 \cdot 10^{-23}$
&  $4.96 \cdot 10^{-23}$ & -
\\
\hline $B_c^+ \to \pi^- \ell^+ \ell^+$ & $1.76 \cdot 10^{-21}$
& $1.75 \cdot 10^{-21}$ &  $3.04 \cdot 10^{-22}$
\\
$B_c^+ \to K^- \ell^+ \ell^+$ &  $1.73 \cdot 10^{-22}$
&  $1.72 \cdot 10^{-22}$ &  $2.89 \cdot 10^{-23}$
\\
$B_c^+ \to D^- \ell^+ \ell^+$ &  $3.20 \cdot 10^{-22}$
&  $3.17 \cdot 10^{-22}$ &  $2.14 \cdot 10^{-23}$
\\
$B_c^+ \to D_s^- \ell^+ \ell^+$ &  $9.17 \cdot 10^{-21}$
 &  $9.10 \cdot 10^{-21}$ &  $5.17 \cdot 10^{-22}$
\\
$B_c^+ \to B^- \ell^+ \ell^+$ & $3.31 \cdot 10^{-28}$
& $2.96 \cdot 10^{-28}$  & -
\\
\hline
\end{tabular*}
%\end{ruledtabular}
\end{table}
%
%
%
%%%%%%%%%%%   END OF TABLE
%
%
The present bounds on the PMNS mixing elements $|U_{N \ell}|$ for
heavy Majorana neutrinos ($m_{N} \geq 100$ GeV) are
 \cite{Nardi}
\begin{equation}
 \sum_N |U_{N e}|^2 \equiv (s_L^{{\nu}_e})^2 \leq 0.005 \ , \qquad
(s_L^{{\nu}_{\mu}})^2 \leq 0.002 \ , \quad (s_L^{{\nu}_{\tau}})^2
\leq 0.010 \ . \label{sLmax1}
\end{equation}
So, Eq.~(\ref{resBr1}) and the bounds in Eq.~(\ref{sLmax1}) allow us to interpret
the ${\cal B}$ values in Table \ref{tabcalA} as upper bound
estimates of the corresponding branching ratios, for the case of
heavy Majorana neutrinos with masses above $100$ GeV,.

As expected, our results in Table~\ref{tabcalA} coincide with
those of Ref.~\cite{Ali}, with discrepancies within 10\% due to variations in the input parameters (a bit larger discrepancies are found in $D^+$ and $D_s$ decays due to the different values we used for $f_D$ and $f_{D_s}$).
The conversion of their theoretical estimates into our notation is
a simple factor $10^{-14}$ due to different units used.
In the cases where Ref.~\cite{Ali} quotes a range, our agreements are with their central values.
The sole exception occurs in $B^+\to\pi^-\tau^+\tau^+$, where their result is almost exactly a factor
10 larger. We can attribute this discrepancy only to a misprint in the power of 10 of their result.

From Table \ref{tabcalA} we see that the
highest upper bounds ($\sim 10^{-20}$) are for the branching ratios
of $B_c^+ \to D_s^- \ell^+ \ell^+$, with $\ell = e$ or
$\mu$. Yet, these bounds are several orders of magnitude too
small to be detected in current and future experiments, like those
at the  LHC-$b$ and Super flavor-factories.
%
%
%
%
%
%We must point out that
%our results shown in Table~\ref{tabcalA} have a remarkable agreement (within
%20\%) with those of Ali $et~al.$ \cite{Ali}, except in $D^+\to
%K^-e^+e^+$, where their result is equivalent to ${\cal B} =
%2.2\times 10^{-23}$ ($i.e.$ 50\% lower) and in $B^+\to
%\pi^-\tau^+\tau^+$, where it is equivalent to ${\cal B} = 0.2 -
%1.2\times 10^{-23}$ ($i.e.$  an order of magnitude larger). The
%conversion of their theoretical estimate to our notation in these
%cases is a simple factor $10^{-14}$ due to different units used.

%%%%%%%%%%%%%%%%%%%%%%%%%%%%%%%%%%%%%%
%%
%%                        OTHER SOURCES of LNV
%%
%%%%%%%%%%%%%%%%%%%%%%%%%%%%%%%%%%%%%%
%
%\section{Other physics sources for the decays $M^+ \to {M'}^-\ell^+\ell^+$}

As a final remark in this subsection, we want to compare these results with what
would be expected if lepton flavor violation came from other sources, namely Supersymmetry with
R-parity violation (RPV), or Left-Right symmetric electroweak theories.
One can anticipate that such LNV processes should involve mass scales well above the
electroweak scale,
typically around the TeV scale, so they can be compared with
the same LNV processes mediated by heavy Majorana neutrinos.
Even though the experimental observation of the LNV meson decays would strongly support
the hypothesis that neutrinos are  Majorana particles, these other sources could produce the
same signals without  involving Majorana neutrinos \emph{directly}, just as it
occurs in neutrinoless double beta decays.

In a supersymmetric extension of the Standard Model that includes RPV, the exchange
of charged lepton or quark superpartners and neutralinos or
gluinos rather than $W$-bosons and Majorana neutrinos can also
induce these LNV meson decays.
%Thus these decays can probe the supersymmetric mass scale and
%constrain the RPV supersymmetric parameters.
%
RPV supersymmetry allows for additional trilinear terms in the
superpotential, of the form:
\begin{eqnarray}
W= \lambda_{ijk}L_i L_j E^c_k +
\lambda^{\prime}_{ijk} L_i Q_j D^c_k + \lambda^{\prime
\prime}_{ijk}U^c_i D^c_j D^c_k,
\end{eqnarray}
where $i,j,k$ denote the families, $L$ and $Q$ denote lepton and
quark iso-doublet chiral superfields and $E^c,U^c$ and $D^c$
charged lepton and quark iso-singlet chiral superfields.
Of these terms, only the second leads to LNV
meson decays \cite{Mohapatra}.
The effective Lagrangian for these decays induced by the RPV terms
can be written as \cite{Hirsch1}
\begin{eqnarray}
L_{eff}^{\Delta L=2}=\frac{G_F^2}{2m_p}\bar{e} (1+\gamma_5)e^c
\left[ \eta_{PS}J_{PS}J_{PS} -\frac{1}{4}\eta_T J^{\mu \nu}_{T}
J_{T\mu \nu} \right],
\end{eqnarray}
where the hadronic currents are
$J_{PS}=\bar{u}^{\alpha}(1+\gamma_5)d_{\alpha}$ and $J^{\mu
\nu}_T= \bar{u}^{\alpha}\sigma^{\mu \nu}(1+\gamma_5)d_{\alpha}$,
with color index $\alpha$ and $\sigma^{\mu
\nu}=(i/2)[\gamma^{\mu},\gamma^{\nu}]$.
Here $m_p$ is the proton mass and the explicit forms of the
parameters $\eta_{PS}$ and $\eta_{T}$ are given in
\cite{eta-forms}.
We should add that, in the case of LNV decays induced by heavy
neutrinos, the effective Lagrangian can also be put in the form
above if we use $\eta_N J^{\mu}_{VA} J_{VA\mu}$, where $J^{\mu
\nu}_{VA}= \bar{u}^{\alpha} \gamma^{\mu}(1-\gamma_5)d_{\alpha}$
and  $\eta_N= \frac{|U_{N \ell}|^2}{10^{-2}} / \frac{100 \ {\rm
GeV}}{m_{N}}$ (see section II.c and Appendix C). If we assume that
the contribution induced by either gluinos or neutralinos is
dominant over the others and the masses of the sfermions are
almost equal, the parameters $\eta_{PS}$ and $\eta_{T}$ are of the
order of
\begin{eqnarray}
\frac{\pi\lambda_{ijk}^{\prime}\lambda_{ij^{\prime}k^{\prime}}^{\prime}}{G^2_{F} m_{\tilde{f}}^4}
 \left(
\frac{\alpha_s m_p}{6 m_{\tilde{g}}} \ + \ \frac{\alpha_2 m_p}{2
m_{\chi}}\right), \label{coee}
\end{eqnarray}
where $\alpha_s, \alpha_2, m_{\chi}, m_{\tilde{g}}$ and
$m_{\tilde{f}}$ denote the strong coupling, $SU(2)$ weak coupling,
neutralino mass, gluino mass and sfermion mass, respectively, and
$\lambda_{ijk}^{\prime}\lambda_{ij^{\prime}k^{\prime}}^{\prime}$ actually depends on the process
($\lambda_{123}^{\prime}\lambda_{111}^{\prime}$ for $B\to D e e$,
$\lambda_{113}^{\prime}\lambda_{112}^{\prime}$ for $B\to K e e$,
$\lambda_{113}^{\prime}\lambda_{111}^{\prime}$ for $B\to\pi e e$,
$\lambda_{122}^{\prime}\lambda_{111}^{\prime}$ for $D\to K e e$,
$\lambda_{121}^{\prime}\lambda_{111}^{\prime}$ for $D\to \pi e e$,
$\lambda_{112}^{\prime}\lambda_{111}^{\prime}$ for $K \to\pi e e$).
Besides $0\nu \beta \beta $ decay, the electron electric dipole
moment experiments lead to the most stringent bounds on single
$\lambda^{\prime}_{111}$, which are $5.5\times 10^{-5}$ for
$m_{\tilde{f}}=100$  GeV and $2.4\times 10^{-7}$ for
$m_{\tilde{f}}=1$ TeV \cite{edm-bound,rpv-bound}.
There are also several bounds on single RPV couplings $\lambda_{11k}^{\prime}$
and $\lambda_{12k}^{\prime}$ coming from experimental
results for forward-backward asymmetries in the fermion pair production reactions
measured at LEP and SLC, and leptonic $\pi$ decays, respectively;
$\lambda_{11k}^{\prime}\lesssim 0.02$
and $\lambda_{12k}^{\prime}\lesssim 0.21$ for $m_{\tilde{d}_{kR}} = 100$ GeV \cite{rpv-bound}.
Imposing those bounds on $\lambda_{111}^{\prime}, \lambda_{11k}^{\prime} (k=2,3)$
and $ \lambda_{12k}^{\prime} (k=1,2,3)$, the magnitudes of the corresponding terms in
Eq.~\eqref{coee} are of order $10^{-8}-10^{-9}$ for
$m_{\tilde{g},\chi,\tilde{f}}=100$ GeV.
These should be compared with the parameter $\eta_N$ for the case
of LNV induced by Majorana neutrinos. For $|U_{N \ell}|^2\sim
10^{-2}$ and $m_N\sim 100$ GeV, the magnitude of $\eta_N$ becomes
unity.
Thus, the above bounds on $\lambda_{ijk}^{\prime}$ would imply
that the RPV supersymmetric contribution to the corresponding LNV decays should be
much smaller than those induced by heavy neutrinos for $m_N=100$
GeV (and maximally allowed $|U_{N \ell}|$).
%, unless
% the hadronic matrix elements in the RPV supersymmetric case were
% unexpectedly large compared to that of the heavy neutrino case, or
Conversely,
the LNV meson
decay experiments can be used to put bounds on the corresponding
$\lambda_{ijk}^{\prime}$ parameters, especially in those cases where the bounds are very loose or still non-existent.

Alternatively, a Left-Right Symmetric Model also involves a large
mass scale which may characterize LNV mediated by heavy  physics
\cite{LRmo02370del}.
%
%Again, since the dynamics of the heavy physics occurs mainly at
%short distances, one may integrate out these heavy degrees of
%freedom, leaving an effective theory of quarks and leptons.
%
In  $SU(2)_L \times SU(2)_R \times U(1)_{B-L}$, this gauge group
breaks down to $SU(2)_L\times U(1)_Y$ via an extended Higgs sector
containing a bi-doublet $\Phi$ and two triplets $\Delta_{L,R}$
whose leptonic couplings generate Majorana neutrino masses and
thus lepton number violation. The $\Delta_{L,R}-$lepton
interactions are not suppressed by lepton masses and have the
structure $L\sim h_{ij}\ \Delta^{++}_{L,R}\ \bar{l^c_i}(1\pm
\gamma_5)l_j+h.c.$, where  the couplings $h_{ij}$ are in general
diagonal and associated with the heavy neutrino mixing matrix. In
this model, short-distance contributions to LNV decays arise from
the exchange of both heavy right-handed Majorana neutrinos and
$\Delta_{L,R}$,  which can be parameterized by \cite{new2},
\begin{eqnarray}
%\frac{\tilde{c_i}}{\Lambda^5}\sim
\frac{g^4_2}{M^4_{W_R}}\frac{1}{M_{\nu_R}},
~~~\frac{g^3_2}{M^3_{W_R}}\frac{h_{ij}}{M_{\Delta}^2}, \label{LR-terms}
\end{eqnarray}
where $g_2, M_{W_R}, M_{\Delta}$ and $M_{\nu_R}$ denote the weak
gauge coupling,  the $SU(2)_R$ gauge boson mass, the triplet
scalar mass and the right-handed neutrino mass, respectively.
These terms are to be compared with $\eta_N G^2_F/m_p$
corresponding to the LNV decays induced by heavy neutrinos.
Imposing the current lower bound of 715 GeV on $M_{W_R}$
\cite{PDG2008} and taking $M_{\nu_R}\sim M_{\Delta} \sim 1$ TeV,
those terms multiplied by $m_p/G^2_F$ are of order of
$10^{-9}-10^{-10}$, which are again very small compared with
$\eta_N^{max}\sim 1$ for $m_N=100$ GeV.

%When we impose the bounds on the RPV couplings assuming the
%supersymmetric mass scale lies between 100 GeV and 1 TeV, the
%contributions associated with the coefficients are less than the
%maximal branching ratios estimated in Table~\ref{tabcalA}. {\bf [IS THAT SO?
%WE SHOULD GIVE A REFERENCE. NOW, READING THE WHOLE SECTION, IT
%SEEMS THAT THIS LAST SENTENCE IS THE REASON FOR THE WHOLE SECTION.
%MAYBE WE SHOULD SAY IT SHORTER].}

%%%%%%%%%%%%%%%%%%%%%%%%%%%%%%%%%%%%%
%
%                         SECTION 4   CONCLUSIONS
%
%%%%%%%%%%%%%%%%%%%%%%%%%%%%%%%%%%%%%

%%%%%%%%%%%%%%%%%%%%%%
%
% Implication for new physics
%
%%%%%%%%%%%%%%%%%%%%%%%%%%

\section{Summary and Conclusions}

%{\bf Would you please add slightly this section?}

We have studied lepton number violating decays of charged $K$,
$D$, $D_s$, $B$ and $B_c$ mesons of the form $M^+\to
{M'}^-\ell^+\ell^+$, induced by the existence of Majorana
neutrinos. These decays violate lepton number by  two units, and
therefore can occur only if neutrinos are of Majorana type. The decays
are sensitive to neutrino masses and lepton mixing, and can also
provide information complementary to neutrinoless double beta
decays. We explore neutrino mass ranges $m_N$ from below 1 eV to
several hundred GeV.

The decay rates are dominated by different weak amplitudes,
depending on the mass of the neutrinos involved in the
intermediate states.

If the mass of the neutrino that dominates the process is below
the mass of the produced meson, we find that the main contribution to the
branching ratio should come from a two-particle intermediate state that
goes on shell, formed by a meson (with the correct flavor) and the
neutrino. These cases have a topology similar to neutrinoless
double beta decay. However, the branching ratios obtained in these
cases are far too small to be detected in foreseen experiments.
Indeed, if the neutrinos involved are standard (masses below 1 eV,
albeit Majorana) we find the branching ratios to be below $10^{-31}$,
and if they are heavier (up to the order of 100 MeV), the
branching ratios to be below $10^{-21}$, which are too small to be
detected in the foreseen future, so we do not go into more refined
calculations in these cases.

Instead, if the neutrino mass is in the range between the masses
of the initial and final meson, the process is dominated by an
intermediate state with just the neutrino, which goes on shell. In
this ``long distance" process, the neutrino is essentially
produced and then it decays. Some of the branching ratios in this
case are now within or near the reach of current or foreseen
experiments, as shown in Table~\ref{tabcalC}. For example,
${\cal B}(K^+\to\pi^-e^+e^+)$ can be up to $10^{-6}$ and
${\cal B}(B_c\to\pi^-e^+e^+)$ up to $10^{-11}$.
Experimental exploration of these decays can then at least provide
upper bounds for the lepton mixing elements of the standard
charged leptons with exotic Majorana neutrinos, bounds that will
be dependent on the mass of the neutrino involved.

Finally, if the process is dominated by a neutrino that is
considerably heavier than the decaying meson, the branching ratio
is again far too suppressed to be experimentally observed, as
shown in Table~\ref{tabcalA} and previously predicted in Ref.~\cite{Ali}.
Indeed, for neutrino masses near 100 GeV or
above, the branching ratios are all below $10^{-20}$. In this case
we also explore other underlying physics sources that could induce
these LNV decays without involving Majorana neutrinos directly,
namely RPV supersymmetric models and left-right symmetric models.
Concerning RPV SUSY, if we impose the current bounds on
the relevant parameters $\lambda_{ijk}^{\prime}$,
the effect of these
interactions on the decays would be lower than the effect of
neutrinos. Otherwise, in general the experimental bound on each of
the decays will impose bounds on its corresponding $\lambda'$
parameter. Finally, concerning left-right symmetric models, their
effect seem to fall far below the contributions of heavy Majorana
neutrinos, and thus these decays may not be useful to put bounds
on those models.
\\

%%%%%%%%%%%%%%%%%%%%%%%%%%%%%%%%%%%%%
%
%                         ACKNOWLEDGEMENTS
%
%%%%%%%%%%%%%%%%%%%%%%%%%%%%%%%%%%%%%

\acknowledgments

G.C.\ and C.D.\ acknowledge support by FONDECYT, Chile, grants
1095196 and 1070227, respectively, and by Anillo Bicentenario,
Chile, grant ACT119. S.K.K. and C.S.K. thank UTFSM for hospitality
and support. S.K.K. work is supported in part by Basic Science
Research Program through the NRF of Korea funded by MOEST
(2009-0090848). C.S.K. work was supported in part by Basic Science
Research Program through the NRF of Korea funded by MOEST
(2009-0088395), in part by KOSEF through the Joint Research
Program (F01-2009-000-10031-0).
\\

%%%%%%%%%%%%%%%%%%%%%%%%%%%%%%%%%%%%%
%
%                         APPENDIX 1
%
%%%%%%%%%%%%%%%%%%%%%%%%%%%%%%%%%%%%%

\newpage

\appendix

\section{The case of Light Neutrinos}

In this appendix we present the calculation of the decay rate for
the process $M^+\to {M'}^- \ell^+\ell^+$ in the case in which
neutrinos are lighter than the mesons in the process. The
calculation is done assuming that the transition matrix element
can be approximated by its absorptive part, which given in
Eq.~\eqref{Mabs}.

After doing the integral in Eq.\eqref{Mabs}, the square of the
amplitude, summed over the final lepton spins, becomes:
\begin{equation}
  |{\cal M}_{abs}|^2   =
  \frac{G_F^4}{16\pi^2} |V_{cb}V_{ud}|^2
     F_{BD}^{+ 2} F_{DD}^{+ 2}
    |U_{N\ell}^2\,  m_N|^2
 \frac{ |{\bf p}_N|^2}{m_{D\ell}^2}  \times {\cal T}  , \label{m2}
\end{equation}
where we have defined ${\cal T}
=
$
 \begin{equation}
  2\,  \textrm{Tr}\left[
 \not\hspace{-2pt}l_2  \left(\not\hspace{-2pt}p_D \not\hspace{-2pt}p_B + m_{D^0}^2
 + E_{D^0}( \gamma^0 \not\hspace{-2pt}p_B +
  \not\hspace{-2pt}p_D  \gamma^0 )\right) \not\hspace{-2pt}l_1
\left(\not\hspace{-2pt}p_B \not\hspace{-2pt}p_D + m_{D^0}^2
 + E_{D^0}(  \not\hspace{-2pt}p_B \gamma^0+
  \gamma^0 \not\hspace{-2pt}p_D
  )\right)\right],\nonumber
\end{equation}
and where $|{\bf p}_N|$ is the neutrino 3-momentum in the rest
frame of the $D$-$N$ pair. Using the well known expression
\begin{equation}
\lambda(x,y,z)\equiv x^2+y^2+z^2-2xy-2yz-2xz ,
\label{lambda1}
\end{equation}
it can be written as $|{\bf p}_N|=
\lambda^{1/2}(m_{D\ell}^2,m_{D_0}^2, m_N^2)/2 m_{D\ell}$.

Since the expression for$ |{\cal M}_{abs}|^2$ in Eq.~\eqref{m2} is
not explicitly covariant, it is convenient to separate the phase
space integral over the $D$-$\ell$-$\ell$ final state
 ($d{\rm ps}_3$) into the 2-body phase spaces for $B\to \ell_1
+X_{D\ell}$ and $X_{D\ell} \to D+\ell_2$, with the invariant mass
of the pair $X_{D\ell}$ integrated over its physical range:
\begin{equation}
\int d\textrm{ps}_3 \equiv \int \prod_{i=1}^3 \frac{d^3
p_i}{(2\pi)^3 2 E_i}(2\pi)^4 \delta^4(\Sigma pi - p_M) =
\int \frac{d m_{D\ell}^2}{2\pi}\     \int
d\textrm{ps}_{(B\to\ell_1\, X_{D\ell})} \int
d\textrm{ps}_{(X_{D\ell}\to D\, \ell_2)},
 \label{lips3}
\end{equation}
where the 2-body phase spaces in their respective rest frames
reduce to:
\[
 d\textrm{ps}_{(B\to\ell_1\, X_{D\ell})}
    = \frac{1}{16\pi^2} \frac{|\tilde{\bf l}_1|}{m_B} d\Omega_{\ell_1}, \quad
 d\textrm{ps}_{(X_{D\ell}\to D\, \ell_2)}
    = \frac{1}{16\pi^2} \frac{|{\bf l}_2|}{m_{D\ell}}
    d\Omega_{\ell_2},
\]
and the 3-momenta in the respective cases are:
\begin{equation}
|\tilde{\bf l}_1| =
\frac{\lambda^{1/2}(m_B^2,m_{D\ell}^2,m_\ell^2)}{2m_B} \
 \quad\textrm{and}\quad |{\bf l}_2| =
\frac{\lambda^{1/2}(m_{D\ell}^2,m_{D}^2,m_\ell^2)}{2m_{D\ell}}.
 \label{l1l2}
\end{equation}
Now, the integration  over $d\Omega_{\ell_2}$ of the non trivial
factor ${\cal T}$ in Eq.~\eqref{m2}
 can be expressed as:
\begin{equation}
\int d\textrm{ps}_{(X_{D\ell}\to D\, \ell_2)}\   {\cal T}   \   =
\frac{1}{16\pi^2}\frac{|{\bf l}_2|}{m_{D\ell}} \
  {4\pi} {\cal R}, \label{intT}
\end{equation}
where ${\cal R}$ is a long expression of dimension $m^6$:
\begin{eqnarray}
 {\cal R} & \equiv &
\Big\{\    8(m_{D^0}^2 + 2 E_{D^0}^2)^2 E_1 E_2 \ +\ 16 (m_{D^0}^2
+ 2 E_{D^0}^2) (E_D E_2 + |{\bf l_2}|^2 )(E_B E_1 - | {\bf
l}_1|^2)\label{Rvalue}
\\
 &+&\ 16(m_{D^0}^2 + 2 E_{D^0}^2) E_{D^0} \Big(\   E_2(E_B E_1 - |
{\bf l}_1|^2) - E_1(E_D E_2+| {\bf l}_2|^2)\   \Big) \nonumber\\
 &+&\ 16 E_{D^0} \Big(\    2E_B(E_B E_1-| {\bf l}_1|^2)(E_DE_2+|
{\bf l}_2|^2) - m_D^2 E_2(E_B E_1 - | {\bf l}_1|^2) - m_B^2 E_1
(E_D E_2 + | {\bf l}_2|^2) \   \Big) \nonumber\\
 &+&\ 8 E_{D^0}^2
\Big(\    E_1 E_2 ( [E_B-E_D]^2 + | {\bf l}_1|^2 + | {\bf l}_2|^2)
 + 2(E_B -E_D)(E_1 | {\bf l}_2|^2 - E_2 | {\bf l}_1|^2) -2 | {\bf l}_1|^2 | {\bf l}_2|^2\   \Big)
\nonumber\\
 &+&\ 8 (m_D^2 E_2 - 2 E_D^2 E_2 - 2 E_D | {\bf l}_2|^2)
 (m_B^2 E_1 - 2 E_B^2 E_1 + 2 E_B | {\bf l}_1|^2 )\ \Big\}
,\nonumber
\end{eqnarray}
where all kinematical variables here are defined in the rest frame
of the $D$-$\ell$ pair and are functions of its invariant mass
$m_{D\ell}$:
\begin{eqnarray}
&& E_D= \frac{m_{D\ell}^2+m_D^2-m_\ell^2}{2 m_{D\ell}},
 E_2= \frac{m_{D\ell}^2-m_D^2+m_\ell^2}{2 m_{D\ell}},
\quad
 E_{D^0}= \frac{m_{D\ell}^2+m_{D^0}^2-m_N^2}{2 m_{D\ell}},
\\
 && E_B= \frac{m_{B}^2+m_{D\ell}^2-m_\ell^2}{2 m_{D\ell}},
 E_1= \frac{m_{B}^2-m_{D\ell}^2-m_\ell^2}{2 m_{D\ell}},
\quad
 | {\bf l}_1| = \frac{\lambda^{1/2}( m_{B}^2, m_{D\ell}^2, m_\ell^2)}{2
 m_{D\ell}}.
 \nonumber
\end{eqnarray}
Finally, since the result in Eq.~\eqref{intT} is independent of
angles, the subsequent integration over $d\Omega_{\ell_1}$ simply
brings a factor $4\pi$.
The decay rate $\Gamma (B^+ \to D^- \ell^+ \ell^-)$ then results
in the expression:
\begin{equation}
\Gamma (B^+ \to D^- \ell^+ \ell^-)
  =
  \frac{G_F^4}{(16\pi^2)^2} |V_{cb}V_{ud}|^2  F_{BD}^{+ 2}
F_{DD}^{+ 2} \frac{|U_{N\ell}^2 m_N|^2}{m_B ^2}
 \int\limits_{(m_D + m_\ell)}\limits^{(m_B-m_\ell)}  \frac{d m_{D\ell} }{2\pi}  \
 \frac{|{\bf p}_N|^2}{m_{D\ell}^2} |{\bf \tilde l_1}|\  |{\bf  l_2}|
 \  \times {\cal R},
 \label{Rate_light}
\end{equation}
where $|{\bf p}_N|$, $|{\bf \tilde l_1}|$, $|{\bf l_2}|$ and
${\cal R}$ were defined above and are explicit functions of
$m_{D\ell}$. The integral in the expression above can be easily
done numerically.

%%%%%%%%%%%%%%%%%%%%%%%%%%%%%%%%%%%%%
%
%                         APPENDIX 2
%
%%%%%%%%%%%%%%%%%%%%%%%%%%%%%%%%%%%%%

\section{The case of intermediate mass neutrinos}

Here we present the meson decay rate $M^+ \to {M'}^- \ell^+
\ell^+$ in the case in which neutrinos have a mass in the
intermediate range $m_{M'} < m_N < m_M$.

The square of ${\cal \tilde M}$ in Eq.~\eqref{amplitude2}  and sum
over external lepton spins of this reduced amplitude, after some
algebra, can be written as :
\begin{equation}
|\tilde{\cal M}|^2 = 32\, m_N^2\ \Big\{
 (m_N^2 - m_{\ell}^2)^2 (l_1\cdot l_2) +
  m_{\ell}^2 \left( (m_N^2 - m_{\ell}^2)^2 - m_M^2 m_{M'}^2 \right)\Big\}
 \label{Majorana}
\end{equation}

The final 3-body phase space can again be separated into two
2-body integrals, and another over the invariant mass of the
intermediate state (which in this case is the neutrino momentum
squared, $p_N^2$):
\begin{equation}
\int d\textrm{ps}_3  =
\int \frac{dp_N^2}{2\pi}\     \int d\textrm{ps}_{(M\to l_1\, N)} \
\int d\textrm{ps}_{(N\to l_2\, M')},
\end{equation}
where $d\textrm{ps}_{(M\to l_1\, N)}$ $= 1/(16\pi^2)( |{\bf \tilde
l}_1|/m_M) d\Omega_1$ and $d\textrm{ps}_{(N\to l_2\, M')}$ $=
1/(16\pi^2) (|{\bf l}_2|/m_N) d\Omega_2$. This time, the
propagator of the intermediate neutrino in the matrix element [see
Eq.~\eqref{amplitude1}] can be approximated by a delta function,
since it is a narrow state:
\[
 \frac{1}{(p_N^2-m_N^2)^2 + m_N^2\Gamma_N^2} \to
\frac{\pi}{m_N \Gamma_N} \delta(p_N^2-m_N^2).
\]
The only term in $|{\cal M}|^2$ that depends on an integration
angle is the one that contains $(l_1\cdot l_2)$ [see
Eq.~\eqref{Majorana}]. In the neutrino rest frame,  $ \int
d\Omega_2 (l_1\cdot l_2) = 4\pi E_1 E_2  $,
 where
 $E_1 = (m_M^2 -m_N^2 - m_\ell^2)/2 m_N$ and
 $E_2 =(m_N^2 + m_\ell^2 -m_{M'}^2)/2 m_N$ are the respective
energies of the external leptons. All other solid angle integrals
give just a factor $4\pi$.
Putting everything together, we obtain the decay rate:
\begin{eqnarray}
\Gamma(M\to M' \ell^+ \ell^+) &=&
\frac{1}{2 m_M} \frac{G_F^4}{4} f_M^2 f_{M'}^2 |{V_{qQ}}{V_{q_2\,
q_1}}|^2  |{U_{N \ell}}|^4
 \int d\textrm{ps}_3\
 \frac{|\tilde {\cal M}|^2}
{(p_N^2-m_N^2)^2 + m_N^2\Gamma_N^2} \label{Interm}  \\
 &=&
\frac{G_F^4}{32\pi^2 m_M}  f_M^2 f_{M'}^2 |{V_{qQ}}{V_{q_2\,
q_1}}|^2  \frac{|{U_{N \ell}}|^4}{m_N \Gamma_N}
 \frac{|\tilde{\bf l}_1|}{m_M}
 \frac{|{\bf l}_2|}{m_N}
\nonumber \\ && \times
 \left\{
 (m_N^2 + m_\ell^2)m_M^2 - (m_N^2 - m_\ell^2)^2 \right\}
 \left\{
 (m_N^2 - m_\ell^2)^2 - (m_N^2 + m_\ell^2) m_{M'}^2 \right\},
\nonumber
\end{eqnarray}
where $|\tilde{\bf l}_1| =
\lambda^{1/2}(m_M^2,m_N^2,m_\ell^2)/2m_M$  and $|{\bf l}_2| =
\lambda^{1/2}(m_N^2,m_{M'}^2,m_\ell^2)/2m_N$.

%%%%%%%%%%%%%%%%%%%%%%%%%%%%%%%%%%%%%
%
%                         APPENDIX 3
%
%%%%%%%%%%%%%%%%%%%%%%%%%%%%%%%%%%%%%

\section{The case of heavy neutrinos}
\label{app:h}

Here we present the derivation of the meson decay rate for
 $M^+ \to M^{\prime -} \ell^+ \ell^+$
($\ell = e, \mu, \tau$) when the exchanged Majorana neutrino is
heavier than the decaying meson. In this case, both weak
amplitudes shown in Fig.~1 contribute with similar strength to the
process.

In order to see the approximations involved, let us first consider
the neutrino mass $m_{N}$ not to be necessarily high. Let us first
recall the valence quark content of the decaying meson $M$ as
$(\bar Q q)$ and of the produced meson $M'$ as $(\bar q_2 q_1)$.
Let us also denote by $J^\mu_{(\bar Q q)} \equiv \bar Q \gamma^\mu
(1-\gamma_5) q$ the weak $V-A$ quark current with flavor change
$q\to Q$. In much the same way as before, we rearrange the lepton
line using charged-conjugated spinors and the Majorana character
of the neutrino, thus appearing an irrelevant Majorana phase
$\lambda_N = \exp(i \delta_N)$. Then the contribution of Fig.~1.a
(and crossed diagram) can be written as:
\begin{eqnarray}
 {\cal M}_{1a} & = & (-1) \ \frac{G_F^2}{2} \
U_{N \ell}^{*2}\, \lambda_N^* \left( V_{q_1 Q}^* V_{q q_2}^*
\right) \langle M' |J^\mu_{(\bar q_2 q)}\  J^\nu_{(\bar Q q_1)}
|M\rangle \label{MFs}\\
 & \times &
  \overline u_{\bar\ell}(l_2) \gamma_{\mu}(1+\gamma_5)
 \left( \frac{\kslash_N+ m_N}{  k_N^2 - m_N^2 + i \Gamma_N m_N  }
+
   \frac{\kslash^{\ '}_N+ m_N}{  {k^{\, '}_N}^2 - m_N^2 + i \Gamma_N m_N
 }\right)
 \gamma_{\nu} (1 - \gamma_5) v_\ell(l_1)
 . \nonumber
\end{eqnarray}
Here we called $k_N$ and $k^{\, '}_N$ the corresponding neutrino
momenta for the two crossings of the external lepton lines. Now,
if the neutrino is heavy, in the denominators we neglect all
except $m_N^2$. In addition, using the orthogonality of the chiral
projectors and the relation $\{\gamma^\mu, \gamma^\nu\} = 2
g^{\mu\nu}$, the expression reduces to:
\begin{equation}
 {\cal M}_{1a} =   \frac{G_F^2}{2} \
U_{N \ell}^{*2}\, \lambda_N^* \left( V_{q_1 Q}^* V_{q q_2}^*
\right) \langle M' | J^\mu_{(\bar q_2 q)}\  J_{(\bar Q q_1)\,\mu}
|M\rangle \ \frac{4}{m_N}
  [\overline u_{\bar\ell}(l_2) (1 - \gamma_5) v_\ell(l_1)]. \label{M1a}
\end{equation}

In much the same way one can treat the contribution of Fig.~1.b,
but now the roles of the flavors $q$ and $q_1$ are interchanged,
thus the quark currents and CKM elements are different:
\begin{equation}
 {\cal M}_{1b} =   \frac{G_F^2}{2} \
U_{N \ell}^{*2} \lambda_N^* \left( V_{q Q}^* V_{q_1 q_2}^* \right)
\langle M' | J^\mu_{(\bar q_2 q_1)}\  J_{(\bar Q q)\,\mu}
|M\rangle \ \frac{4}{m_N}
  [\overline u_{\bar\ell}(l_2) (1 - \gamma_5) v_\ell(l_1)]. \label{M1b}
\end{equation}
The largest uncertainty here is in the determination of the
hadronic matrix element. As a first estimate, reasonable for the
level of accuracy we seek, we can separate the currents inserting
the vacuum and assuming it saturates the expression (vacuum
saturation approximation). We must then do a Fierz rearrangement
in $M_{1a}$ to match the flavors, which then mismatches the color,
inducing a suppression factor $1/N_c$. The hadronic currents then
reduce to their corresponding decay constants, $f_M$ and $f_{M'}$,
and the total amplitude for heavy neutrino exchange (``h'')
becomes:
\begin{eqnarray}
{\cal M}_{\rm h} &=&
 {\cal M}_{1a}+ {\cal M}_{1b}
 \nonumber\\
 &=&   \frac{G_F^2}{2} \
U_{N \ell}^{*2} \lambda_N^*  \left[V_{q Q}^* V_{q_1 q_2}^* +
\frac{V_{q_1 Q}^* V_{q q_2}^*}{N_c}\right]
 f_M f_{M'} (p_M \cdot p_{M'})
 \ \frac{4}{m_N}
  [\overline u_{\bar\ell}(l_2) (1 - \gamma_5) v_\ell(l_1)]. \label{M1ab}
\end{eqnarray}
One may neglect the $1/N_c$ term, except if the other term is much
more CKM-suppressed. The square and sum over final polarizations
of this amplitude is:
\begin{equation}
 |{\cal M}_{\rm h}|^2 = |{\cal K}_{\rm h}|^2\
\, 32 \,(p_M \cdot p_{M'})^2 \ (l_1 \cdot l_2) , \label{M2heavy}
\end{equation}
where we have gathered all constant factors under the symbol
\[ |{\cal K}_{\rm h}|^2 =
  G_F^4 \left|\frac{U_{N \ell}^{*2}}{m_N}\right|^2
\left| V_{q Q}^* V_{q_1 q_2}^* + \frac{V_{q_1 Q}^* V_{q
q_2}^*}{N_c}  \right|^2  f_M^2 f_{M'}^2 .
\]
The decay width $\Gamma(M^+ \to M^{\prime -} \ell^+ \ell^+)$ can
now be calculated explicitly, by integrating $|{\cal M}_{\rm
h}|^2$  over the final phase space. This time we express the
$M'$-$\ell$-$\ell$ phase space as:
\[
\int d{\rm ps}_3 = \int \frac{d m_{\ell\ell}^2}{2\pi }
\int d{\rm ps}_{(M\to M'\, X_{\ell\ell} )}
\int d{\rm ps}_{(X_{\ell\ell}\to \ell\ell)}
\]
where $d{\rm ps}_{(M\to M'\, X_{\ell\ell} )} =(1/16\pi^2)(|{\bf
p}_{M'}|/m_M)d\Omega_{M'}$ and $d{\rm ps}_{(X_{\ell\ell}\to \ell\,
\ell )} =(1/16\pi^2)(|{\bf l}_2|/m_{\ell\ell})d\Omega_{\ell_2}$.
In the frame of the lepton pair, the integrand \eqref{M2heavy} is
independent of angles, so the integral over $d\Omega_{\ell_2}$ is
a simple factor of $4\pi$, and equally for the subsequent integral
over $d\Omega_{M'}$.

The decay rate is then:
\begin{equation}
 \Gamma(M^+ \to M^{\prime -} \ell^+ \ell^+)
  =
  \frac{1}{2!}\frac{1}{8\pi^2} \frac{|{\cal K}_{\rm h}|^2 }{m_M}
\int \limits_{4 m_\ell^2}\limits^{(m_M-m_{M'})^2} \frac{d
m_{\ell\ell}^2}{2\pi} \frac{|{\bf p}_{M'}|}{m_M} \frac{|{\bf
l}_2|}{m_{\ell\ell}} (m_M^2+m_{M'}^2-m_{\ell\ell}^2)^2\,
(m_{\ell\ell}^2 - 2 m_\ell^2), \label{G2}
\end{equation}
where $|{\bf p}_{M^{\prime}}|$
 $= \lambda^{1/2}( m_M^2,m_{M'}^2,m_{\ell\ell}^2)/2 m_M$
  and $|{\bf l}_2| $
 $=\lambda^{1/2}(m_{\ell\ell}^2, m_\ell^2, m_\ell^2)/2 m_{\ell\ell}$.
The factor $1/2!$ appears because there are two identical
particles in the final state. This integral can be easily done
numerically.

\newpage
%%%%%%%%%%%%%%%%%%%%%%%%%%%%%%%%%%%%%
%
%                        REFERENCES
%
%%%%%%%%%%%%%%%%%%%%%%%%%%%%%%%%%%%%%


\begin{thebibliography}{10}
\bibitem{double_beta}
A similar -- but much more complex -- problem of proliferation of
intermediate states occurs in nuclear $\beta\beta 0\nu$ decay,
where a multitude of nuclear and hadronic excitations should be
considered in the intermediate state.
\end{thebibliography}
\end{document}